\documentclass[a4paper,12pt]{article}
\usepackage{rotating}
\usepackage[utf8]{inputenc}
\usepackage[english]{babel}
\usepackage{amsmath}
\usepackage{authblk}
\usepackage{graphicx}
\usepackage[singlespacing]{setspace}
\usepackage[headheight=1in,margin=1in]{geometry}
\usepackage{fancyhdr}
\usepackage{lipsum}
\usepackage{hyperref}
\usepackage{xcolor}
\usepackage{amsfonts}
\usepackage{amsmath}
\usepackage{latexsym}
\usepackage{caption}
\usepackage[labelfont=bf]{caption}
\usepackage{makecell}
\usepackage{url}
\usepackage{subcaption}
\usepackage{tabularx}

\usepackage{caption}
\usepackage{natbib}
\setcitestyle{authoryear,open={(},close={)}} 

\date{}

\begin{document}
\title{Origins of City Reputation: \\Measuring Trust and Bias in Online Reviews Using Agentic AI and NLP}
\author{
Rafiazka Hilman,\textsuperscript{1,2}, Júlia Koltai\textsuperscript{1,3}
}

\maketitle
\begin{center}
{
1. MTA–TK Lendület “Momentum” Digital Social Science Research Group for Social Stratification, ELTE Centre for Social Sciences, Budapest, Hungary\\
2. Informatics Institute, University of Amsterdam, Amsterdam, The Netherlands\\
3. Institute of Empirical Studies, ELTE Faculty of Social Sciences, Budapest, Hungary
}\\
\end{center}

\vspace{10mm}

\begin{abstract}
Online reviews aggregate millions of individual judgments into signals that shape where people eat, stay
and travel to, and increasingly read them as evidence of the reputation of places. Nevertheless, whether such
reputations really reflect the places or more the instruments used to measure them is unclear. Here we compare two
instruments for measuring trust and bias in 100{,}000 Yelp reviews from ten US\ cities: a large language model based agentic system including an evaluator critic loop, and a corpus-based tool built on pretrained classifiers, lexicons
and rule-based features. Our results show that the two instruments agree on credibility, but not at all on
plausibility. They also respond to different use of  words and phrases. The corpus-based better flagged complaints and
generalizations, whereas the agent better quoted references to social groups. We also found that with five examples, the agent treats ordinary superlatives as bias far more often than with none or one. Aggregated to cities, both
instruments find little variation as less than half a percent of the variance of review level reputation
lies between cities. This small differences between cities mostly arise from different levels of trust (in case of the agentic tool) and from bias (in case of the corpus-based tool), and the two instruments rank the cities differently. Reputation derived from platform reviews
therefore depends on how the reviews are processed and analyzed. As platforms delegate this to
language models, the definitions and examples built into these models become part of how places are
valued.\\


\end{abstract}

\clearpage

\section{Introduction}
\label{Introduction}
Before choosing a restaurant, a hotel or a city to visit, most people now read what strangers have written
about it. Reviews shape which products and places consumers choose \citep{babicrosario2016, babicrosario2020}. They find reviews posted in online communities especially credible for experience
goods, whose quality can only be judged after consumption \citep{bae2011}. Aggregated over many consumers, this reliance on reviews has measurable economic consequences for businesses. On Yelp, a one-star increase in a restaurant's average rating raises its revenue by 5-9\%, and so far the
platform has partly replaced older forms of reputation such as chain affiliation \citep{luca2016}. Review
platforms therefore do more than record opinions as they aggregate thousands of individual judgments into
signals that steer where people eat, stay and travel.

These signals are also attached to travel destinations. Tourism study treats online travel reviews as a form of electronic
word-of-mouth that reveals the image visitors form of a destination after their trip
\citep{guo2021, marineroig2020}. Reviews express how safe, attractive or authentic a place appears to
those who wrote them, and they transmit that image to prospective visitors \citep{marineroig2020}. The process, which influences which
image emerges and which sinks depends partly on the platform since the same destination can be represented differently on
different review sites \citep{guo2021}. Reviews can also carry social prejudice into the image of a place.
For example, in Brooklyn, Yelp reviewers described restaurants in a predominantly Black neighborhood with words such as
\textit{ghetto} and \textit{sketchy}, while praising a predominantly White neighborhood as cozy and
authentic, a pattern \citet{zukin2017} call "discursive redlining". Aggregated across many reviews, such
language can shape the reputation of a whole city or district.

Whether review content builds a credible reputation of a place depends on two properties: how far the reviews
can be trusted, and how far they are biased. Trust matters because a reputation built from reviews works only if
readers believe them. Consumers and travelers adopt review information they find credible and discount the
rest \citep{erkan2016, ismagilova2020}. Trust cannot be taken for granted, and its absence
is not random noise that cancels out in averages. About 16\% of restaurant reviews on Yelp are filtered as
suspicious, and businesses are more likely to commit review fraud when their reputation is weak or competition
is strong \citep{luca2016fraud}. Firms also post promotional reviews of themselves and negative reviews of
competitors where incentives favor it \citep{mayzlin2014}, so review sites must verify reviews to maintain
users' trust \citep{zelenka2021}. Even genuine ratings are a selective sample, because people mostly write reviews after very good or very bad experiences, which makes the average star rating a biased estimate of quality \citep{schoenmueller2020}. How far a review can be trusted must therefore be judged from its text, not only from its
rating.

Bias matters because reviews comment on more than the business. They also describe the people and places around
it, and prejudice in that language travels from individual writers to the image of a whole area
\citep{zukin2017}. Bias ranges from overt insults to subtle generalizations about groups of people or places.
The two properties vary independently namely a detailed, credible account of a meal can still stereotype the
neighborhood around the restaurant, and a review free of bias can be fabricated. A measure of place reputation
therefore has to assess both, and weigh each review by how trustworthy and how unbiased it is.

Computational studies measure trust and biases mostly with corpus-based natural
language processing (NLP) such as sentiment classifiers \citep{hutto2014, loureiro2022}, lexicons of biased language \citep{recasens2013},
inference models for detecting inconsistency \citep{williams2018, laurer2024}, and embedding-based association tests that reveal
human-like biases in language \citep{caliskan2017}. These instruments are transparent, because each score can be traced back to a fixed rule or model output,
and scalable, because they can be applied to millions of reviews at low cost. However, they detect whether
certain cues are present in a text, not what these cues mean in their context. The same word can carry
very different meanings. \textit{Never} marks a sweeping generalization in \textit{"people from this area never
pay"}, but an ordinary complaint in \textit{"the food never arrived"}, and \textit{"not rude at all"} contains the very
word a lexicon counts as negative. Conversely, prejudice is often expressed without any flagged word, through
framing such as \textit{"you know what this part of town is like"}. Because word choice differs between groups of
writers, these errors are systematic rather than random (Section~\ref{Measuring trust and bias in reviews with NLP}),
and they do not cancel out when scores are aggregated to places. A reputation built on such counts may
therefore reflect how people write more than what they report. Furthermore, many
NLP measures of bias in rest on loosely defined constructs \citep{blodgett2020}.

Large language models offer a different kind of instrument for the measurement of trust and biases. They can annotate text at a quality that
matches or exceeds crowd workers on many tasks \citep{gilardi2023}, although their reliability varies
widely between tasks and constructs \citep{ziems2024}. Configured as agents, they can also explain their
judgments, quote the evidence for them, and revise them in response to feedback from a second model
\citep{madaan2023, wang2024}. Such agentic assessments have two further advantages. They apply a construct defined in words, through
the same definitions and coding scheme that a human coder would use, rather than through a fixed set of proxy
cues. Language models can annotate text from the same codebook instructions given to human coders
\citep{gilardi2023}. In addition, every score is tied to quoted passages from the review, so each judgment
can be checked against the text it is based on. Grounding model outputs in quoted evidence allows such
checks to be made directly \citep{deyoung2020, gao2023}. It is unclear, however, whether they measure the same theoretical constructs of trust and biases as corpus-based methods, or different
ones. The answer matters beyond methodology and has epistemological significance: if the two approaches disagreed, the reputation of a city would depend on the instrument used to measure it, not only on what reviewers wrote.

To test this, we compare agentic and corpus-based measurements of trust and bias in 100{,}000 Yelp reviews from ten
US\ cities, and trace how each of these measures shapes city level reputation. Biases are operationalized separately for implicit and explicit biases, while trust is operationalized by credibility and plausibility. An agent built on Gemma 4 31B assesses
these implicit and explicit bias, credibility and plausibility through an evaluator critic loop, prompted
with zero, one or five examples. For the other tool, a corpus-based measures the same dimensions with
pretrained classifiers, lexicons and rule-based features. Both tools report scores on the same five point scale from 0 to 1 for each review.
We combine trust and bias into a review level reputation signal, whose city mean defines city
reputation. We are seeking answers to two research questions. First, to what extent do agentic assessments of bias and
trust align with corpus-based measurements (RQ1)? Second, how do the two approaches shape our
understanding of city reputation as an aggregate outcome of online reviews
(RQ2)?

\subsection{Reviews as an infrastructure of valuation}
\label{Reviews as an infrastructure of valuation}

Online reviews belong to the reputation systems that let strangers trust one another in markets without
personal acquaintance \citep{resnick2000}. A rating summarizes past experiences and makes it visible to future
buyers, and its effects on choices and sales are well documented \citep{chevalier2006, luca2016, babicrosario2016}. Platform
studies add that reviews are not neutral records of quality. Review sites decide which voices count, how
they are aggregated and how the result is displayed. In doing so they reorganize how hotels,
restaurants and destinations are valued \citep{orlikowski2014}. This process belongs to the wider
platformization of social life, in which platforms become the infrastructure through which public value is
produced and contested \citep{vandijck2018}. Algorithms that rank, filter and score content are part of this
infrastructure since they shape what counts as relevant knowledge out of all that has been reported
\citep{gillespie2014}.

Seen this way, any measure computed from reviews is itself an act of valuation. When platforms
or city marketers score reviews by their credibility or bias and aggregate these scores, they add a further layer
to the platform's apparatus. The choice of instrument is therefore not a technical detail. It determines
which reviews count as trustworthy and which places appear reputable.

\subsection{From review content to the image and characteristics of travel locations}
\label{From review content to the image and characteristics of travel locations}

Reviews also describe travel destinations. Tourism study uses online travel reviews to reconstruct the image visitors
form of a destination, including its perceived safety \citep{marineroig2020}, and shows that this image
depends on the platform on which reviews appear \citep{guo2021}. Place branding study treats such travel destination images
as dynamic outcomes of many online conversations rather than of official marketing alone
\citep{kavaratzis2013}.

Nevertheless, reviews do not only construct an image of a destination, but they can also inform us about real social changes of an area. Urban study has used the same data in the measurement of social changes within the city, for instance Yelp
listings predict neighborhood change \citep{glaeser2018}, and the restaurants people visit reveal how
segregated urban consumption is \citep{davis2019}. Location-based platforms likewise show that people mostly
visit places of their own socioeconomic class, so digital traces of consumption carry information on the city's social
divisions \citep{moro2021, hilman2022}. The language of these reviews can also contribute to the understanding of a city or district social characteristics. \citet{zukin2017} found that Yelp reviewers framed
restaurants in a predominantly Black Brooklyn neighborhood through danger and dirt, and those in a
predominantly White one through authenticity that may steer investment. Reviews
can thus transmit prejudice from individual writers to the image of a whole area. Whether they do so at the
scale of cities, and whether automated instruments can detect such language reliably, remains open.

\subsection{Measuring trust and bias in reviews with NLP}
\label{Measuring trust and bias in reviews with NLP}

For trust, studies on review credibility frequently identifies textual cues that readers use to decide whom to believe, which can be the detailedness or
the balance of the text, or the absence of exaggeration \citep{filieri2016}. In such decisions readers often rely on heuristics rather than
careful scrutiny when they judge online information \citep{metzger2010}, and platforms face deliberate
manipulation through promotional and fake reviews \citep{luca2016fraud, he2022, zelenka2021}. Computational approaches detect deceptive reviews from linguistic features \citep{ott2011} and estimate
credibility from proxies such as specificity, length and narration in the first person. These proxies only
partly match human judgment. Some correspond to the cues readers use, as longer and more detailed reviews
are perceived as more helpful \citep{hong2017}. Yet people are poor judges of deception. In a controlled
study, human judges identified fabricated hotel reviews at close to chance level and tended to accept reviews
as genuine, while a classifier trained on linguistic features distinguished them reliably \citep{ott2011}.
The same study found that fabricated reviews used the first person more often than genuine ones, so a cue
that intuitively signals experience at first hand can also signal deception. Computational proxies may
therefore encode the heuristics people use, including their errors, and agreement with human intuition alone
does not establish that a measure of credibility is valid.

For bias, corpus-based NLP offers lexicons of biased language \citep{recasens2013}, toxicity classifiers,
and embedding-based association tests that reveal stereotypes encoded in language
\citep{caliskan2017, garg2018}. These instruments hold transparency and scalability, but they carry biases of
their own. Toxicity classifiers flag dialects and identity terms disproportionately \citep{davidson2019, sap2019},
and bias can enter NLP through data, annotation, representations, models and research design
\citep{hovy2021}. A critical survey found that most NLP work on bias leaves the underlying construct
undefined \citep{blodgett2020}. Measurement theory frames the problem precisely: an operationalization is
valid only if it measures the construct it claims to measure \citep{jacobs2021}, and text-as-data methods
must be validated rather than assumed to work \citep{grimmer2022}.

\subsection{Measuring trust and bias in reviews with language models and agents}
\label{Measuring trust and bias in reviews with language models and agents}

Large language models can raise the quality of automated text analysis. They annotate
text at or above the quality of crowd workers on many tasks \citep{gilardi2023}, detect sentiment and
offensiveness far more accurately than dictionary-based text analysis \citep{rathje2024}, can reproduce the response
patterns of human subpopulations \citep{argyle2023}, and consequently, are increasingly proposed as tools for the social sciences
\citep{bail2024}. Their performance, however, varies widely between tasks, and they perform worst on
constructs that require social or contextual interpretation \citep{ziems2024}. Their outputs are also
sensitive to how they are prompted, including the choice, order and formatting of few-shot examples, which can shift results
substantially \citep{zhao2021, lu2022, sclar2024}.

Nevertheless, agentic designs can add explanation to the LLM's choices and revise them if they are needed. An evaluator model can quote the evidence for its judgment, and a
second model can audit and refine it \citep{madaan2023, wang2024}. According to previous studies, models used as judges agree well with human
preferences in some settings \citep{zheng2023}, but self-correction without external feedback often fails to
improve reasoning, and it succeeds mainly when reliable external feedback is available
\citep{huang2024, kamoi2024}. Whether such agents measure trust and bias more validly than corpus-based
instruments, or simply differently, has not been tested yet on the same reviews.

Based on the overview of existing studies, two gaps follow that we explore. First, agentic and corpus-based measurements of trust and bias have not been compared on
identical texts with a shared scale, and accordingly it is unknown whether they capture the same constructs (RQ1). Second, studies that infer the reputation of places from online reviews treat city reputation as an
aggregate outcome of these reviews, but rarely ask how the measurement approach shapes this outcome, and
therefore whether different approaches lead to a different understanding of the same city's reputation (RQ2). The complete pipeline is provided in Supplementary Material A.

\section{Data and Methods}
\label{Data and Methods}

\subsection{Data}
\label{Data}

We analyze restaurant and business reviews from the Yelp Open Dataset \citep{yelpdataset} for the ten cities with the most reviews in the US\
comprising Indianapolis, Nashville, New Orleans, Philadelphia, Reno, Saint Louis, Saint Petersburg, Santa
Barbara, Tampa and Tucson. From each city file we take a random sample of 10{,}000 reviews after
discarding reviews with empty text and duplicated review identifiers, giving 100{,}000 reviews in total. The same reviews are analyzed in each measurement. Each review is assigned to the city of the file it comes from. The measurement
instruments receive only the review text, while star ratings, vote counts and business attributes are used
later, as covariates in the reputation model.

\subsection{Agentic measurement}
\label{Agentic measurement}

The agents are built on Gemma 4 with 31 billion parameters~\citep{gemma2026}, accessed through Ollama. Two agents, one for bias and one for trust, assess each review separately. Each consists of an evaluator and a critic that use the same model, the same definitions and the same coding scheme. The coding scheme consists of conceptual definitions and level descriptions provided to the agent. It assigns
every score to one of five levels of an ordinal scale (0, 0.25, 0.50, 0.75, 1), and describes in words what
each level means, so that the model, like a human coder, scores a review by matching it to the closest
description.

\paragraph{Constructs and coding scheme.} The bias agent scores \textit{implicit bias} (subtle preferential framing,
stereotypes or assumptions that are not stated directly), \textit{explicit bias} (words that openly and
unfairly favor or disfavor individuals, groups or entities) and an overall bias score $b_i$. Its
definitions state explicitly that a strong opinion tied to a described experience (\textit{"the soup was cold"})
is not bias. Bias is a statement about people or places that is unfair, unsupported or generalized, such as stereotypes about staff,
customers, neighborhoods or the city. The trust agent scores \textit{credibility} (claims grounded in
concrete, potentially verifiable details such as dishes, prices, waiting times or staff actions),
\textit{plausibility} (internal consistency and realism of the account) and an overall trust score $t_i$.
Its definitions state that trust concerns the credibility of the account, not whether the reviewer liked the
business, so a negative review can be fully trustworthy. For bias, the levels range from no bias (0, a neutral or experience based account), through a single loaded
word (0.25), clearly one sided framing or an unsupported generalization (0.50) and strong loaded language or
stereotyping of people or places (0.75), to pervasive prejudice that dominates the review (1). For trust, they
range from not credible (0, contradictory, implausible or purely promotional), through mostly vague claims
(0.25), some concrete details mixed with generic claims (0.50) and a specific, coherent account at first hand
(0.75), to a detailed, consistent account grounded in observable facts (1). The same scale applies to the
two dimension scores and to the overall score of each agent. The prompts used to design both the trust and bias agents in this study are provided in Supplementary Material C.

\paragraph{Evaluator.} The evaluator receives the definitions, the coding scheme, step-by-step instructions and the
review text, and returns a structured assessment. It first extracts exact quotations from the text as evidence (biased words or phrases for bias, and supporting evidence or red flags for trust), then writes a justification of at most three sentences that refers to the quotes, and only then assigns each score by choosing the level of the coding scheme that best
matches the review. This evidence-first order ties the score to text at each review so the reader can inspect each score against the text. The instructions direct the agent to use only the review text and to
make no assumptions about the business, its rating or its popularity.

\paragraph{Grounding tool and critic.} An automated tool verifies that all quotations appear exactly as written in the original review, after normalizing Unicode, quotation marks, whitespace and letter case. The critic receives the
review, the current assessment and this grounding report. It checks whether each quote fits the definition, whether clear evidence was missed, and whether each score matches the description of its level in the coding scheme, given the quoted evidence. For both agents, the definitions state that the tone of a review is not a criterion: a strong opinion tied
to a described experience is not bias, and a negative review can be fully trustworthy. The trust critic
additionally checks that the review's sentiment did not drive the trust score, because readers tend to judge
negative reviews as more credible than positive ones \citep{lo2019, chen2013}, while unspecific enthusiasm
raises suspicion of promotional writing \citep{filieri2016}. A trust score driven by tone would turn the
reputation measure into a measure of sentiment. The critic either accepts the assessment or revises it. It may revise only if it names a specific problem
that can be traced to the review text, such as a quote that does not appear in the review, a quote that
does not fit the definition, clear evidence that the evaluator missed, or a score whose level does not match
the quoted evidence. Otherwise acceptance is the default, and the critic is instructed not to change a score
merely because it was asked to review it, since language models asked to check their own answers tend to
revise answers that were already correct \citep{huang2024}. The critic runs for at most three rounds and stops when it accepts or when no score changes by 0.05 or higher are made. Each round's scores are stored, so the evaluator's initial scores remain available for comparison.

\paragraph{Prompt designs.} We run each agent in three designs namely zero-shot, one-shot and five-shot. The examples are synthetic reviews written and scored by a human, covering all five scale levels. The one-shot example is the first of the five, so the designs are nested. The examples are shown to both evaluator and critic, with the instruction to judge each review on its own merits rather than copy the examples' scores.

\paragraph{Inference settings.} Outputs are constrained to a JSON schema during decoding. Because a language
model writes scores as text, it could still return values outside the scale, such as 1.5 or 75. Each reply is
therefore validated against the same schema before it is used. The validation checks that all required fields
are present (the quotes, the justification, the scores and, for the critic, the verdict), that each field has
the expected type, that every score lies between 0 and 1, and that the verdict is either accept or revise. A
reply that fails any of these checks is discarded rather than corrected. Scores within the range but between
two scale levels are rounded to the nearest level when methods are compared. The model also has to repeat the
identifier of the review, and any mismatch is recorded.

We use a temperature of 0.2, which keeps sampling close to deterministic, and fix the random seed of every
call so that each output can be reproduced. The value of the seed has no substantive meaning. We fixed it in
advance at 42, a common convention, for every review rather than choosing among several runs. Each call
receives its own seed, 42 for the evaluator and $42 + 1000k$ for critic round $k$, and a repeated request
adds one to the seed of the failed attempt. The offset of 1,000 between rounds ensures that no two calls for
the same review share a seed, so a critic round never repeats the random draws of the evaluator or of an
earlier round. The context window is set to 16{,}384 tokens, enough to hold the longest prompt, the critic
prompt with five examples and a long review, and the extended reasoning mode of the model is switched off,
because the evaluator already states its evidence and justification before scoring.

A request is repeated when its reply is invalid or cut off at the maximum output length, or when the server
returns an error or does not respond in time, in which case the request waits between 5 and 120 seconds
before it is repeated. After five failed attempts, the review is left unscored. Because the prompt combines
fixed instructions, definitions, examples and, for the critic, the current assessment with the review itself,
a very long review could make the prompt exceed the context window. In that case, Ollama silently drops part
of the prompt rather than returning an error. We therefore check the prompt length reported for every request
and stop the run whenever a prompt fills 98\% of the window or more, so that no review is scored on truncated
text. Each review is scored once per design. The formal expression used in agentic measurement is provided in Supplementary Material B.

\subsection{Corpus-based measurement}
\label{Corpus-based measurement}
The corpus-based tool measures the same dimensions with fixed pretrained models, lexicons and rule-based
features. It is independent of the agentic measurement at every step. It receives only the review text, and
uses none of the agents' prompts, examples, quotes, justifications or scores. None of its models or
thresholds was trained, tuned or adjusted on the agents' output, and its mapping rules were fixed before the
two measurements were compared. The two tools share only the sample of reviews and the five point scale on
which they report their scores. Reviews are split into sentences with pySBD \citep{sadvilkar2020}, processed with spaCy \citep{honnibal2020} for tokens, lemmas and named entities. Each pretrained model is applied to single sentences, or to sentence pairs for the inference model, rather
than to whole reviews. These models accept at most 512 tokens per input (384 for the sentence embedding
model), because their representation of word positions is defined only up to that length, and longer inputs
cause an error rather than a weaker result. Longer sentences are therefore truncated to this limit, and only
the first 60 sentences of a review are processed. Because the limits apply to single sentences, they bind
only for unusually long text, such as a review without punctuation that the segmenter returns as a single
block, or a review of more than 60 sentences. In such reviews, cues in the omitted text are not counted, so
the corpus-based scores of these reviews may understate the cues they contain.

\paragraph{Trust.} Each cue operationalizes part of the definitions given to the agents. Because
credibility is defined as grounding in concrete, potentially verifiable details, its core cue counts such
details, namely numeric details (prices, times, quantities, dates) and named entities. Readers judge detailed
reviews as more credible \citep{filieri2016}, and genuine reviews describe concrete circumstances more
specifically than fabricated ones \citep{ott2011}. Review length enters because longer reviews have more room
for such detail and are perceived as more helpful \citep{hong2017}. The use of the first person marks an
account of personal experience, as the definition requires, although it is an ambiguous cue, since fabricated
reviews use it more often than genuine ones \citep{ott2011}. Exaggeration, the rate of exclamation marks and
all-caps words, lowers credibility, following evidence that readers distrust exaggerated reviews
\citep{filieri2016}. Plausibility is the internal consistency of the review. Whether the described visit took
place cannot be verified from the text, but whether the review contradicts itself can. We measure it as one
minus the largest contradiction probability between pairs of its sentences, according to a DeBERTa-v3 natural
language inference model \citep{laurer2024}, since a single contradiction is enough to undermine an account.
It is computed on all sentence pairs, or on a reproducible random subset of 45 pairs for longer reviews,
because the number of pairs grows quadratically with the number of sentences (45 pairs correspond to a
review of ten sentences). Plausibility is undefined for single-sentence reviews.

\paragraph{Bias.} As for trust, we chose each cue to capture one element of the definitions given to the agents, rather than
general negativity, because a strong opinion tied to a described experience is not bias. Explicit bias, words
that openly and unfairly favor or disfavor people or groups, is measured with the sentence level identity
attack and insult probabilities of the unbiased-toxic-roberta classifier \citep{detoxify2020}. We use these
two of its labels rather than overall toxicity, because they target demeaning language about people and
their group membership, whereas profanity or anger about a meal is not bias. The classifier was trained to
reduce false alarms on mentions of identity groups \citep{borkan2019}, a known weakness of such tools
\citep{davidson2019, sap2019}. The bias lexicon of \citet{recasens2013}, derived from edits that removed
partial wording from Wikipedia, captures one sided framing that toxicity classifiers miss. Implicit
bias, preferential framing or stereotypes that are not stated directly, draws on two cues. The first is the
rate of generalization markers (e.g.\ \textit{always}, \textit{never}, \textit{typical}, \textit{these
people}), because describing people in abstract and general terms rather than through specific acts is a
linguistic signature of stereotyping \citep{semin1988, maass1999}. The second is the largest absolute
projection of a sentence that mentions people, groups or places onto a pleasant-unpleasant valence axis in
sentence-embedding space, built from the WEAT attribute words, which captures evaluative associations
attached to social targets \citep{caliskan2017}. The third is a single-category embedding association test
(SC-WEAT) that measures how much more a sentence resembles generalizing statements (\textit{"People around here are
always rude"}) than specific, experience-based ones (\textit{"Our server forgot to bring our drinks for twenty
minutes"}), using all-mpnet-base-v2 sentence embeddings \citep{reimers2019}. Unlike word lists, it also
recognizes generalizations phrased in words no list contains. We did not use a supervised bias classifier,
because no labeled data for bias in reviews exist that would match these definitions.

\paragraph{Mapping the two measurements.} To compare the scores of the corpus-based evaluation with the agentic measurement, we map the NLP cues onto the same five point scale with rules fixed before any comparison with the agent (Table~\ref{tab:rules}). Overall bias is the maximum of the implicit and explicit levels, so a review is as biased as its strongest form. Overall trust is the mean of credibility and plausibility. All thresholds are reported so that the mapping can be reproduced and varied. The technical notes on the corpus-based measurement are provided in Supplementary Material D.

\begin{table*}[t]
  \centering
  \small
  \caption{\textbf{Apriori rules mapping corpus-based cues onto the five point scale.} Rules are
  applied from the top. The first that applies sets the level, otherwise the level is 0. Counts are per review. Toxicity values are maximum sentence level probabilities. SC-WEAT values are maximum sentence level effect sizes. "Mentions people" means at least one sentence refers to people, groups or places. Heavy exaggeration ($\geq 5$ exclamation marks or all-caps words per 100 words) lowers both trust dimensions by one level.}
  \label{tab:rules}
  \begin{tabular}{@{}l p{0.4\textwidth} p{0.4\textwidth}@{}}
    \hline
    Level & Explicit bias & Implicit bias \\
    \hline
    1.00 & identity attack $\geq 0.8$, or identity attack and insult both $\geq 0.5$ &
           $\geq 3$ generalization markers, SC-WEAT $\geq 0.8$ and mentions people \\
    0.75 & identity attack $\geq 0.5$, or insult $\geq 0.8$ & SC-WEAT $\geq 0.8$ and mentions people \\
    0.50 & insult $\geq 0.5$, or $\geq 3$ bias-lexicon words &
           SC-WEAT $\geq 0.5$, or $\geq 2$ markers and mentions people \\
    0.25 & insult or identity attack $\geq 0.2$, or $\geq 1$ bias-lexicon word & $\geq 1$ marker, or SC-WEAT $\geq 0.2$ \\
    \hline
    Level & Credibility & Plausibility (largest contradiction probability) \\
    \hline
    1.00 & $\geq 4$ concrete details, first-person and $\geq 80$ words & $< 0.20$ \\
    0.75 & $\geq 2$ concrete details, first-person and $\geq 40$ words & $< 0.50$ \\
    0.50 & $\geq 1$ concrete detail, or first-person and $\geq 40$ words & $< 0.80$ \\
    0.25 & first-person, or $\geq 20$ words & $< 0.95$ \\
    \hline
  \end{tabular}
\end{table*}

\subsection{Reputation}
\label{Reputation}
Following the conceptual model of the study, a review contributes to the reputation of its city in
proportion to how trustworthy and how unbiased it is. We define the review level reputation signal as
$R_i = t_i (1 - b_i)$ and city reputation as its mean over the $N_c$ reviews of city $c$:

\begin{equation}
  R_c = \frac{1}{N_c} \sum_{i \in c} t_i\,(1 - b_i).
  \label{eq:reputation}
\end{equation}

We compute $R_c$ separately for each method, that is, for the three agent designs and the corpus-based
measurement, using the overall scores of trust and bias.

\subsection{Statistical analysis of the Research Questions}
\label{Statistical analysis of the Research Questions}

Because all methods score the same reviews, we can compare them with paired statistics and report effect sizes throughout, which is important as With about 10{,}000 reviews per city, almost any difference are significant on conventional significance.

\paragraph{Agreement between methods (RQ1).} We measure review level agreement with Spearman's rank
correlation $\rho$ and with quadratic-weighted Cohen's $\kappa$ \citep{cohen1968} on scale levels, rounding  scores to the nearest level and we interpret $\kappa$ following \citet{landis1977}. To test whether the two dimensions of each construct relate to each other in the same way across methods, we
compare the two scores of each review, implicit with explicit bias and credibility with plausibility, using
the Vargha--Delaney $A$ statistic \citep{vargha2000}. $A$ is the share of reviews in which the first score is
higher than the second, where reviews with equal scores count as one half. Equal scores are common because
the scores take only five levels, and counting them as one half makes $A = 0.5$ mean that neither score tends
to be higher. Differences are tested with the Wilcoxon signed-rank test. Differences in mean scores between all methods are tested with the Friedman test, with Kendall's $W$ as effect size, followed by pairwise Wilcoxon signed-rank tests with Holm correction \citep{holm1979}. To show how often the two approaches disagree strongly, we report the share of reviews on which the agent and
the corpus-based measurement differ by at least two levels of the five point scale.

\paragraph{Language behind the judgments.} To better understand the origins of potential similarities and differences between the models' assessments, we explore whether they rely on different linguistic features in their assessments. We compare the most frequent terms of those texts, which the models based their assessments on. We compare these frequencies with the weighted log-odds ratio with an informative Dirichlet prior \citep{monroe2008}. For the agentic models, the target text consists of the passages it quoted as evidence, and the comparison text is the
full text of all reviews it scored, including the reviews from which the quotes were taken. A term is thus
distinctive if the agent quotes it more often than its frequency in the reviews would lead one to expect.
Because the quotes are also part of the comparison text, this contrast is slightly conservative. For the
corpus-based model, which quotes nothing, the target text consists of the reviews it scored high (bias
$\geq 0.5$; trust $\geq 0.75$) or low (trust $\leq 0.25$), and the comparison text consists of its other
reviews. For deriving the linguistic features from the text we count words and two-word phrases within sentences, exclude contractions, function words and words shorter than three letters, keep generalization markers such as \textit{always} and \textit{never}, and require at least five occurrences.

\paragraph{City differences and reputation (RQ2).} Differences between cities in each score are tested with
the Kruskal--Wallis test, with $\eta^2_H = (H - k + 1)/(n - k)$ as effect size \citep{tomczak2014}, and
summarized by city rankings, whose agreement across methods we measure with Spearman's $\rho$.

To identify whether differences in city reputation arise from trust or from bias, we decompose each city's
deviation from the mean reputation of all reviews exactly into a trust part, a bias part and a remainder
\begin{equation}
  R_c - \bar R = \underbrace{(\bar t_c - \bar t)(1 - \bar b)}_{\text{trust}}
  \;\underbrace{-\;\bar t\,(\bar b_c - \bar b)}_{\text{bias}} \;+\; \text{remainder},
  \label{eq:decomposition}
\end{equation}
where $\bar R$, $\bar t$ and $\bar b$ are the means of reputation, trust and bias over all reviews, $\bar t_c$
and $\bar b_c$ are the means of trust and bias over the reviews of city $c$, and the remainder collects the
interaction and within-city covariance terms. The decomposition follows the logic of decomposing a difference
between two rates into components \citep{kitagawa1955, dasgupta1993}. Each part shows how much the city's
deviation in one factor would change its reputation if the other factor were at the overall mean. Holding
the other factor at the overall mean is one of several possible reference choices, but it matters little
here, because the interaction term is the product of two small city deviations. We summarize the
decomposition by the share of the absolute city deviations attributable to each part (Supplementary Material E).

To separate city effects from review characteristics, for each method we fit a linear mixed model of $R_i$
with a random intercept per city and five standardized review level predictors, namely star rating, the
logarithms of one plus the useful, funny and cool vote counts, and the logarithm of one plus the review
length in words. Vote counts and review length are strongly right skewed. Most reviews receive no or few
votes and are of moderate length, while a small number receive many votes or are very long. Without the
logarithm, these few extreme reviews would dominate the estimated coefficients and their standardized
contributions. The logarithm also reflects the diminishing importance of each additional unit, since the
difference between no vote and one vote, or between 20 and 120 words, says more about a review than the
difference between 100 and 101 votes, or between 1,020 and 1,120 words \citep{gelman2020}. One is added
before taking the logarithm because the counts include zero, for which the logarithm is undefined. Models are
estimated by restricted maximum likelihood with several optimizers, keeping the solution with the highest
finite likelihood. Ten cities are few for a multilevel model, and rules of thumb often call for at least 20
to 30 higher level units \citep{bryan2016}. These rules concern inference on effects of higher level
predictors and on the variance between groups, whose standard errors are too small when groups are few
\citep{maas2005}. The coefficients and variance components themselves are estimated without bias in linear
models with few clusters when restricted maximum likelihood is used \citep{elff2021, mcneish2016}. We
therefore use the model only to estimate the share of variance between cities and the effects of review
level predictors, and we do not test the variance between cities. Cluster robust standard errors would not
avoid the problem, because they are too small with few clusters \citep{cameron2008} and do not estimate the
variance between cities. The Kruskal--Wallis effect sizes, which require no assumption about city effects,
provide an independent check on the size of city differences. We report the intraclass correlation (the
share of variance between cities) and the marginal and conditional $R^2$ of \citet{nakagawa2013}.
Contributions of the predictors are expressed as SHAP values, which equal $\beta_j (x_{ij} - \bar x_j)$
\citep{lundberg2017}, and are averaged by city to describe each city's profile. To explore whether city effects relate to characteristics of the cities, we correlate the estimated city
effects with the logarithms of population, area and number of restaurants and with restaurant density, using
Spearman's $\rho$ with Holm correction. As the study includes only ten cities, we treat these associations as
exploratory.

\section{Results}
\label{Results}

We compare four measurements of the same 100{,}000 Yelp reviews from ten US\ cities, a corpus-based NLP and the Gemma 4 31B agent prompted with zero, one or five  examples. All four report bias
and trust on a shared five point scale from 0 to 1. Because every method scores the same reviews, we compare them
with paired statistics and report effect sizes rather than $p$-values as with 10{,}000 reviews per city, almost every difference would be significant on the conventional significance levels.
 
\subsection{Agentic and corpus-based measures assess the same reviews differently}
\label{Agentic and corpus-based measures evaluate the same reviews differently}
 
\begin{figure*}[t]
  \centering
  \includegraphics[width=\textwidth]{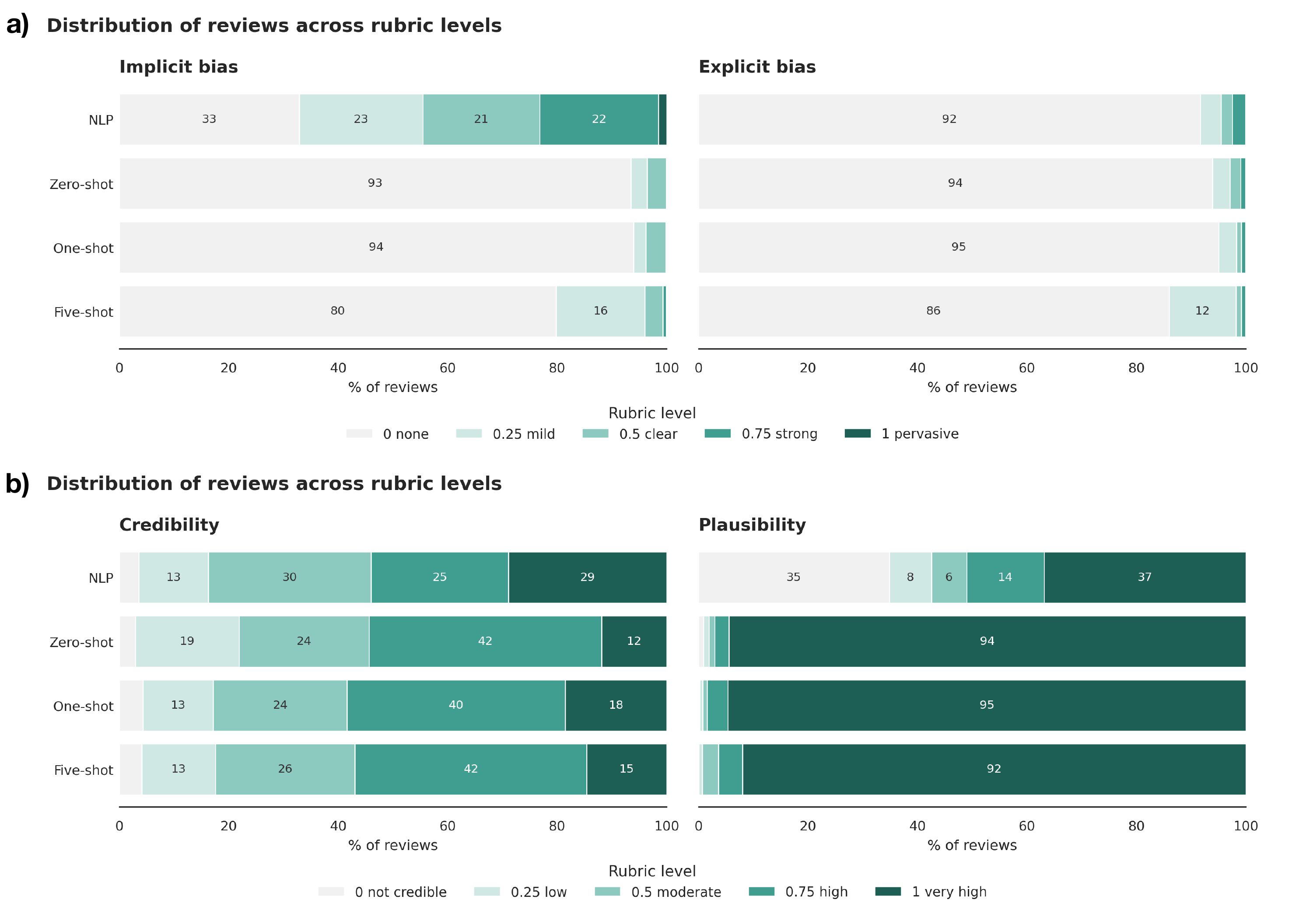}
  \caption{\textbf{The agent and the NLP use the same scale in sharply different ways.}
  Share of reviews at each scale level (0 to 1 in steps of 0.25) for the NLP based model and for Gemma 4 31B
  with zero, one and five examples.
  (a) Bias. The NLP assigns implicit bias to two-thirds of reviews (33\% at score 0), whereas the
  agent finds none in 93-94\% of reviews (zero- and one-shot) and in 80\% with five examples; explicit bias
  is rare for every method (86-95\% at score 0).
  (b) Trust. Both approaches spread credibility across the scale, but plausibility diverges: the agent rates
  92-95\% of reviews as very plausible, whereas the NLP based model places 35\% at the lowest level.
  Numbers inside bars are percentages; segments below 6\% are unlabeled.}
  \label{fig1}
\end{figure*}

The two approaches disagree first about how much trust and bias the reviews contain (Figure~\ref{fig1}). The NLP based model finds implicit bias in two-thirds of reviews with only 33\% having
score 0, and 22\% reach score 0.75, the level of strong bias. The agent without examples finds none in 93\% of the reviews, which is 94\% with one example (Figure~\ref{fig1}a). Within the same review, the NLP based model is more likely to find implicit bias above explicit bias in most cases (Vargha-Delaney $A = 0.81$), whereas the agent rates the two at almost the same low level ($A = 0.50$-$0.54$). Accordingly, while there is a difference in the estimation of implicit bias between the NLP based and the agentic method, explicit bias, by contrast, is rare for every method, with 86-95\% of reviews at score 0. Moreover, the examples shift the agent's calibration. With five examples, the share of reviews without implicit bias falls from 93\% to 80\%, and 16\% move to the mild level. We return to this effect below, because it changes not only how much bias the agent finds but also what it counts as a bias.

Trust shows a different pattern. Both approaches spread credibility across the whole scale, with most reviews
between scores 0.50 and 1 (Figure~\ref{fig1}b). Plausibility, however, separates the two approaches sharply.
The agent gives 92--95\% of reviews score 1, whereas the NLP based model places 35\% at score 0. Within the
same review, the comparison of the two dimensions mirrors the one for bias. The NLP based model rates
credibility above plausibility about as often as below it (Vargha--Delaney $A = 0.53$), whereas the agent
rates plausibility above credibility in almost every review ($A = 0.08$--$0.11$). For the agent, plausibility
thus sits at the ceiling and the variation in trust comes almost entirely from credibility, while for the NLP
based model both dimensions vary. This suggests that the two approaches measure different constructs under
the name of plausibility rather than two estimates of the same construct. The difference lies in the
definitions. The NLP based model measures only logical consistency, as one minus the largest contradiction
probability between pairs of sentences (Section~\ref{Corpus-based measurement}), whereas the agent's
definition also covers whether the account flows logically and whether its claims are realistic
(Supplementary Material C). The NLP based model finds at least one pair of sentences with a contradiction
probability of 0.95 or higher, which corresponds to score 0 (Table~\ref{tab:rules}), in 35\% of reviews,
whereas the agent finds a plausibility problem in only 5--8\%. Whether the contradictions detected by the
inference model reflect real inconsistencies or false alarms cannot be decided without human coding.

Differences in mean scores confirm this pattern. The methods differ strongly in implicit bias (Friedman test,
Kendall's $W = 0.52$), overall bias ($W = 0.52$) and plausibility ($W = 0.51$), but hardly in explicit bias
($W = 0.03$), credibility ($W = 0.03$) and overall trust ($W = 0.07$). The largest differences lie between the
corpus-based tool and the agent, whose mean implicit bias is 0.341 and 0.025--0.04 respectively, and whose
mean plausibility is 0.526 and up to 0.981. Because of the large sample, nearly all pairwise differences remain
significant after Holm correction, even where $W$ is negligible (Supplementary Material G).

\pagebreak
 
\subsection{The two approaches agree on credibility, but not on bias or plausibility}
\label{The two approaches agree on credibility, but not on bias or plausibility}
 
\begin{figure*}[t]
  \centering
  \includegraphics[width=\textwidth]{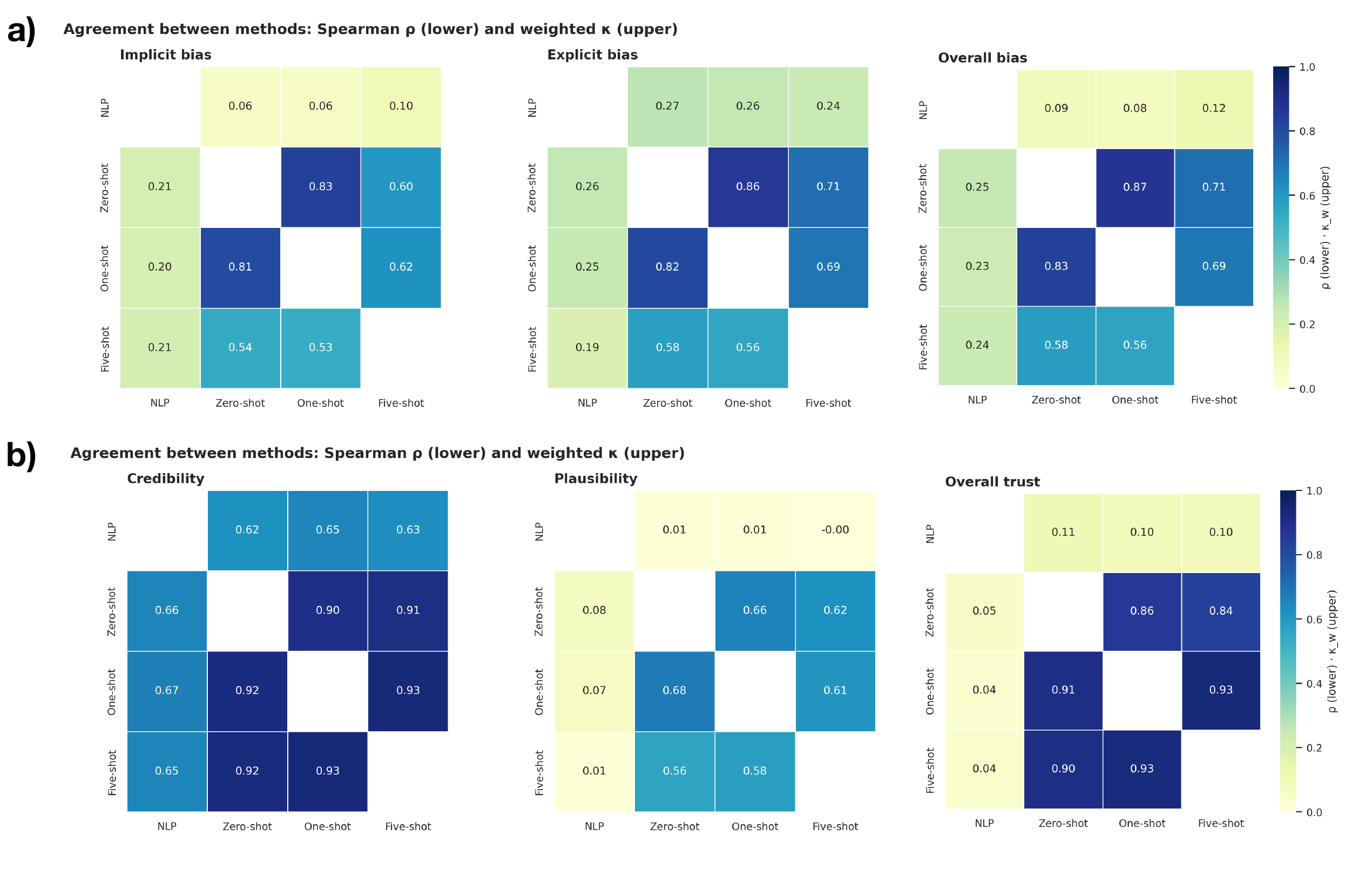}
  \caption{\textbf{Agentic and corpus-based measures agree on credibility, but not on bias or plausibility.}
  Review level agreement between all pairs of methods: Spearman's $\rho$ (lower triangle) and
  quadratic-weighted Cohen's $\kappa$ on scale levels (upper triangle).
  (a) Bias. Agreement between the corpus-based tool and the agent is weak for implicit bias
  ($\rho = 0.20$--$0.21$, $\kappa = 0.06$--$0.10$), explicit bias ($\rho = 0.19$--$0.26$,
  $\kappa = 0.24$--$0.27$) and overall bias ($\rho = 0.23$--$0.25$, $\kappa = 0.08$--$0.12$).
  (b) Trust. The two approaches agree substantially on credibility ($\rho = 0.65$-$0.67$,
  $\kappa = 0.62$-$0.65$) but not on plausibility ($\rho \leq 0.08$), so agreement on overall trust stays negligible ($\rho = 0.04$-$0.05$).
  The three agent designs agree closely with one another (trust $\rho = 0.90$-$0.93$; bias
  $\rho = 0.53$-$0.83$), so the divergence lies between measurement approaches, not between prompts.}
  \label{fig2}
\end{figure*}

At the level of individual reviews, agreement depends on the dimension measured (Figure~\ref{fig2}).
For bias, the NLP based model and the agent rank reviews only weakly alike (implicit bias: Spearman
$\rho = 0.20$--$0.21$; explicit bias: $\rho = 0.19$--$0.26$; overall bias: $\rho = 0.23$--$0.25$). They also
rarely assign the same score (quadratic weighted $\kappa = 0.06$--$0.10$ for implicit bias, $0.24$--$0.27$ for
explicit bias and $0.08$--$0.12$ for overall bias). Agreement on scores is somewhat higher for explicit bias,
because both approaches give most reviews score 0, but it remains weak. For credibility, agreement is substantial ($\rho = 0.65$-$0.67$, $\kappa = 0.62$-$0.65$). Nevertheless, for plausibility, it is absent ($\rho \leq 0.08$). As overall trust averages the two dimensions, the plausibility gap drives agreement on overall trust down to $\rho = 0.04$-$0.05$, masking the strong agreement on credibility.
 
The disagreement between the NLP based and the agentic models lies between measurement approaches, not between prompts. The three agent designs agree closely with one another for the overall trust ($\rho = 0.90$-$0.93$) and moderately to strongly for overall bias
($\rho = 0.56$-$0.83$), with five-shot the least aligned. Large disagreements are common. In 38-41\% of reviews, the NLP based model estimates overall bias at least two scale levels above the agent, while the reverse occurs in less than 1\%. For trust, the direction flips: the agent rates 16-29\% of reviews at least two levels higher than the NLP based model, and the NLP base one does so for 12\%.
 
These results answer the first research question in two parts. Agentic and corpus-based measurements align where trust rests on observable features of the text, such as concrete detail. They diverge where the judgment requires interpretation, as in whether a generalization is unwarranted, or whether an experience is
believable.
 
\subsection{Each approach responds to different language}
\label{Each approach responds to different language}
 
\begin{figure*}[t]
  \centering
  \includegraphics[width=\textwidth]{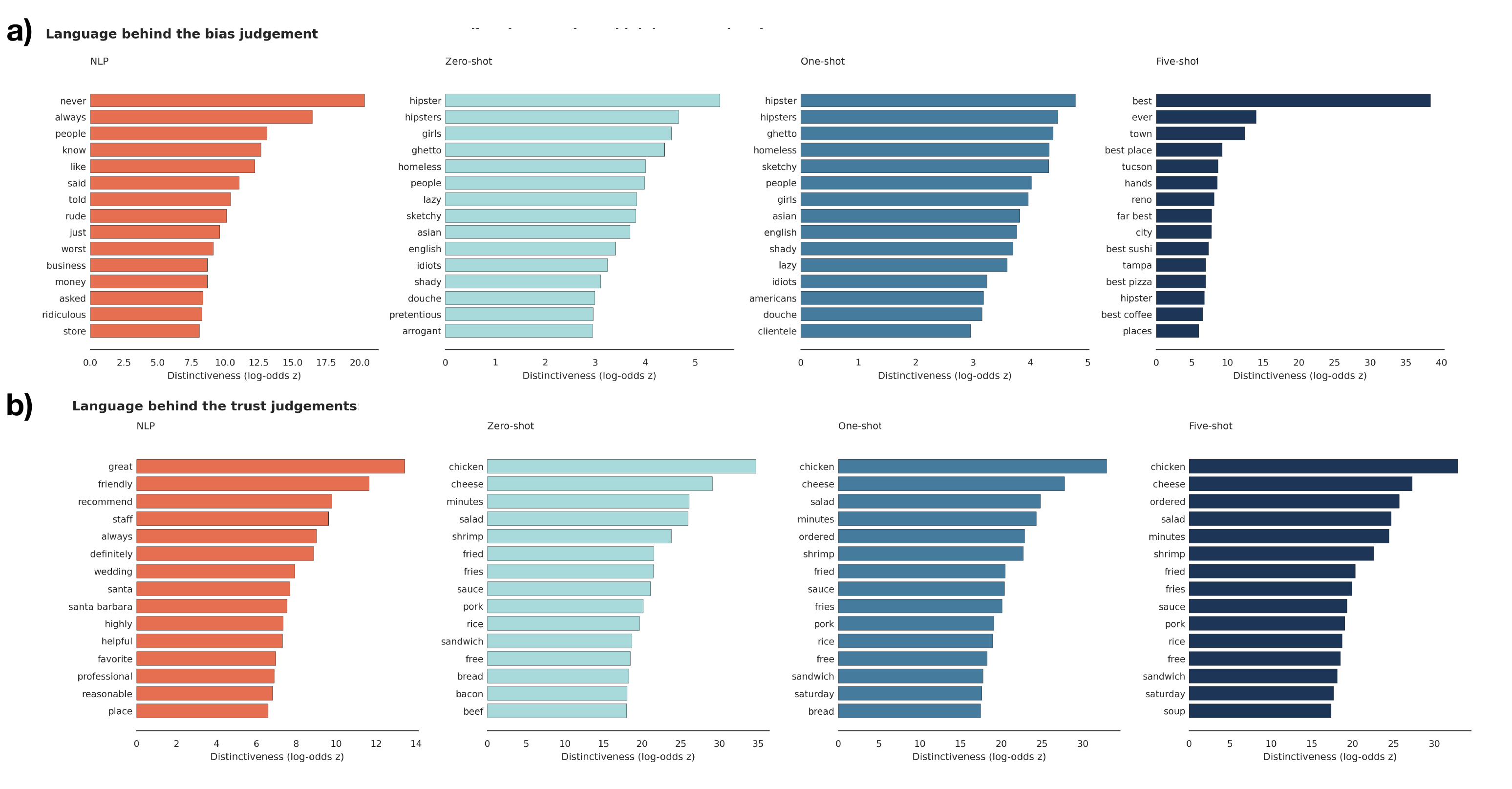}
  \caption{\textbf{The two approaches attend to different language.}
  Terms most distinctive of each method's evaluation, ranked by the weighted log-odds $z$-score within an
  informative Dirichlet prior \citep{monroe2008}: higher $z$ means stronger over-representation. For the
  agent, the target texts are the passages it quoted; for the NLP based method, which quotes nothing, these are the reviews it scored high, compared with other reviews.
  (a) Bias. The NLP based method responds to generalizing and complaint words (\textit{never, always, rude,
  worst}), whereas the zero- and one-shot agent quotes references to social groups and places
  (\textit{hipster, ghetto, homeless}). With five examples, the agent flags superlatives
  (\textit{best, ever, best place}) and quotes biased wording in 21\% of reviews instead of the 7-8\% of the lower shot models, a sign
  that the examples can shift what it treats as bias.
  (b) Trust. The agentic models grounds trust in concrete, checkable details (dishes, prices, waiting times), whereas
  the NLP based method associates high trust with generic praise (\textit{great, friendly, recommend}), words
  the agents quote as red flags.
  Contractions, function words and terms shorter than three letters are excluded; generalization markers
  (\textit{always, never}) are retained.}
  \label{fig3}
\end{figure*}

To understand why different models estimate the plausibility and biases differently, we analyzed the linguistic features of the reviews that characterize the models' decisions. We found that the divergence has a lexical signature (Figure~\ref{fig3}). To identify the terms most distinctive of each method's judgments we compared the language a method attends to with the language of all reviews \citep{monroe2008}. For the agent, this is the text quoted as evidence, while for the NLP based method (which quotes nothing) it is the reviews it scored high. For the comparison we used weighted log-odds $z$-score values, as previously explained in Section~\ref{Statistical analysis of the Research Questions}).
 
According to the results, for bias, the NLP based model is sensitive to generalizing and complaint language. Its high-bias reviews are marked
by \textit{never} ($z \approx 20$), \textit{always}, \textit{rude}, \textit{worst} and \textit{ridiculous}
(Figure~\ref{fig3}a). Many of these are ordinary expressions of dissatisfaction rather than biased
statements about people. The zero- and one-shot agent instead quotes references to social groups and
places such as \textit{hipster}, \textit{ghetto}, \textit{homeless}, \textit{girls} and nationality terms. It
flags such wording in only 7-8\% of reviews, consistent with its low prevalence estimates. The agent thus
locates bias in statements about who people are, whereas the NLP locates it in how strongly
reviewers complain. Interestingly, five examples change this picture. The five-shot agent quotes biased wording in 21\% of reviews, and
its most distinctive term becomes \textit{best} ($z \approx 38$), followed by \textit{ever}, \textit{best place} and city names. The coding scheme leaves open whether such superlatives count as bias. It defines bias as unfair, unsupported
or generalized evaluation of individuals, groups or entities, which the zero and one-shot agents applied
mainly to statements about people and places. One of the five examples, however, scores an unsupported
superlative about a restaurant (\textit{"the best pizza in the city, nothing else even comes close"}) as mild bias
(Supplementary Material C). The five-shot agent appears to have generalized this single case to ordinary
praise, a prompt design effect that any few-shot measurement of bias should test for.
 
For trust, the two approaches rely on opposite kinds of terms (Figure~\ref{fig3}b). The agentic models
ground trust in concrete, checkable details. In 95\% of all reviews, every agent design quotes at least one passage as supporting evidence, on average
about 3.5 passages per review. In the remaining reviews it quotes none, as the instructions allow when a
review contains no concrete detail. Compared with the reviews as a whole, the most distinctive terms of these
quotes are dishes (\textit{chicken}, \textit{cheese}, \textit{shrimp}) and specifics such as waiting times
(\textit{minutes}). The NLP based model associates high trust with generic praise, among others \textit{great},
\textit{friendly}, \textit{recommend} and \textit{definitely}. In contrast, the agent quotes precisely this vocabulary (\textit{great}, \textit{best} and \textit{amazing}) as red flags. What one method reads as a sign of a satisfied, credible customer, the other reads as a sign of promotional or inauthentic writing.
 
\pagebreak
 
\subsection{Cities barely differ in bias and differ only slightly in trust}
\label{Cities barely differ in bias and differ only slightly in trust}
 
\begin{figure*}[tbp]
  \centering
  \includegraphics[width=\textwidth]{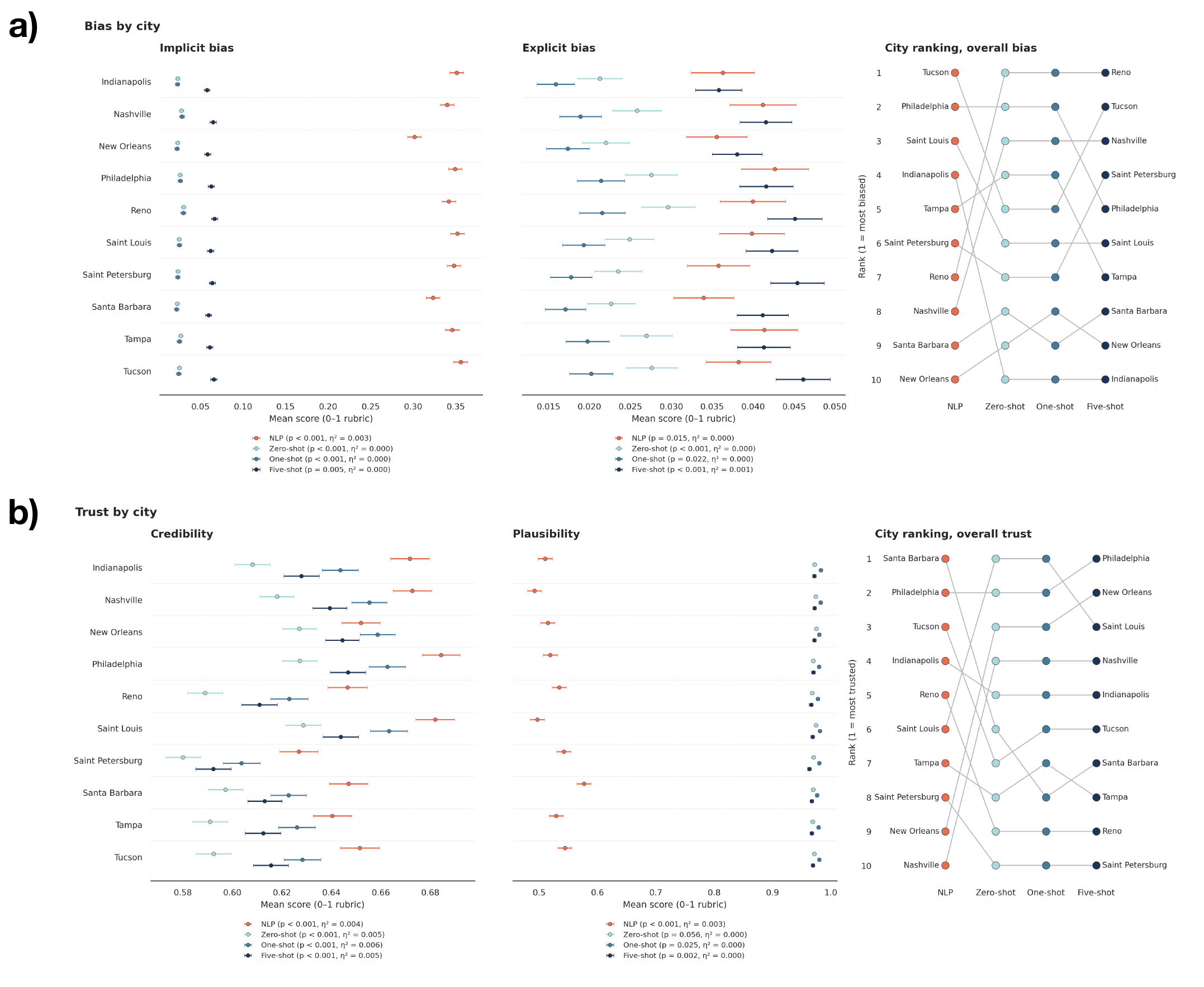}
\caption{\textbf{Cities barely differ in bias and differ only slightly in trust, and the methods rank them
  differently.}
  City means with 95\% confidence intervals (left and center) and city rankings by overall score (right;
  lines connect the same city across methods).
  (a) Bias. City membership explains almost none of the variation (Kruskal--Wallis
  $\eta^2_H \leq 0.003$ for every method), so the rankings largely reflect noise.
  (b) Trust. Average credibility shows a small but consistent pattern across the agent designs
  ($\eta^2_H = 0.005$--$0.006$), with Saint Petersburg lowest and Philadelphia, New Orleans and Saint Louis
  highest. Average plausibility does not differ between cities for the agent ($\eta^2_H = 0.000$). The NLP
  based method finds city differences of a similar, very small size for credibility ($\eta^2_H = 0.004$) and
  plausibility ($\eta^2_H = 0.003$), but orders the cities differently. Santa Barbara ranks first in overall
  trust for the NLP based method but seventh to eighth for the agent.
  Rank~1 = most biased (a) or most trusted (b).}
  \label{fig4}
\end{figure*}

The second research question asks how these measurements shape city reputation. The first overall answer is that
city differences are small for every method (Figure~\ref{fig4}). For the average level of bias, city membership explains
almost none of the variation, reflected by Kruskal-Wallis $\eta^2_H \leq 0.003$ for the NLP based method and
$\eta^2_H \leq 0.001$ for every agent design (Figure~\ref{fig4}a). The city rankings by bias therefore
largely reflect noise, and they differ across methods accordingly. This suggests that average bias does not differ between cities under either method.
 
Trust is where cities differ at all. The average level of credibility shows a small but consistent pattern
across the agent designs ($\eta^2_H = 0.005$--$0.006$; Figure~\ref{fig4}b). Saint Petersburg has the lowest
credibility in every design, and Philadelphia, New Orleans and Saint Louis rank among the highest. With about
10{,}000 reviews per city, the standard error of a city mean is roughly 0.003, so a pattern that repeats across
three independently prompted designs is unlikely to be noise. Yet the city level accounts for less than one
percent of the variance. Plausibility does not differ between cities according to the agent
($\eta^2_H = 0.000$), because nearly all reviews receive the highest score. The NLP based method finds city
differences of a similar, very small size for credibility ($\eta^2_H = 0.004$) and, unlike the agent, also for
plausibility ($\eta^2_H = 0.003$), yet it orders the cities differently. Santa Barbara ranks first in overall
trust for the NLP based method but seventh to eighth for the agent, and New Orleans moves from the upper half
for the agent to ninth for the NLP based method. Because the differences between cities are so small under
both instruments, small differences in what each instrument measures are enough to reorder the cities. The
city ranking a study reports thus depends on the instrument it uses.
 
\clearpage
 
\subsection{City reputation differences arise through different mechanisms}
\label{City reputation differences arise through different mechanisms}
\begin{figure*}[tbp]
  \centering
  \includegraphics[width=\textwidth]{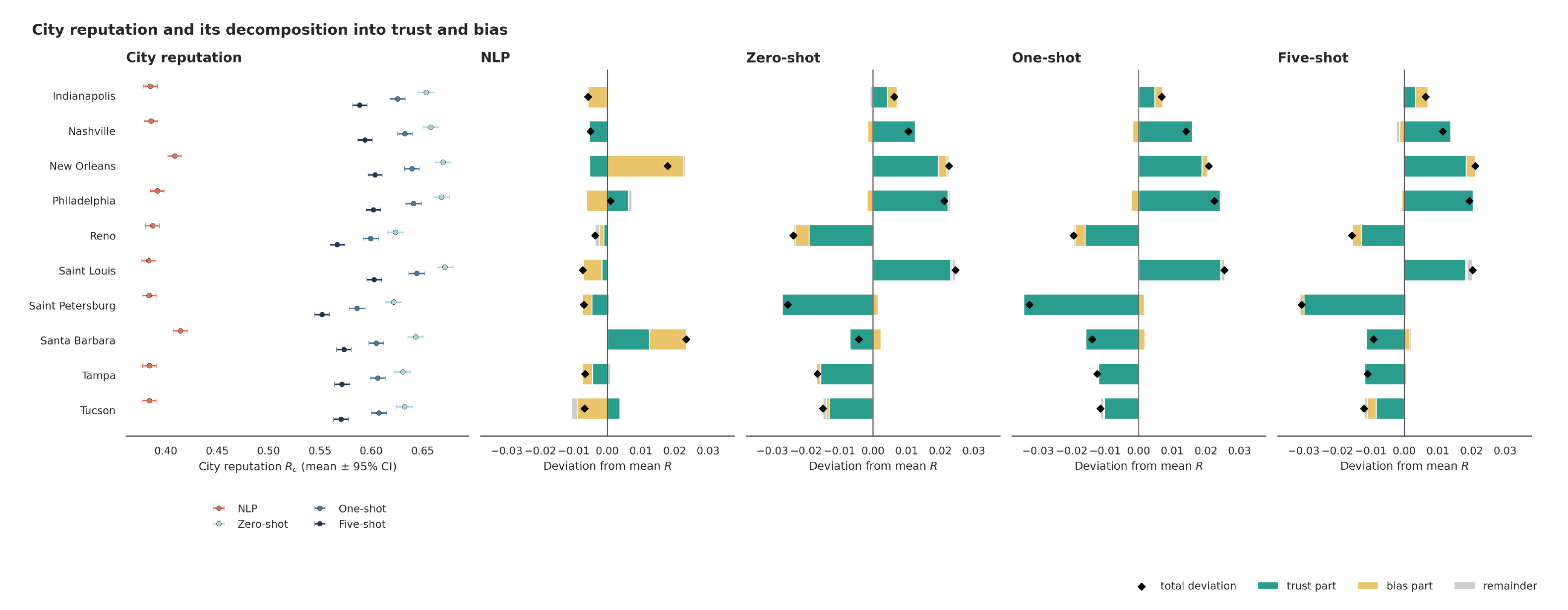}
  \caption{\textbf{Small differences in reputation arise from trust for the agent and from bias for the
  NLP.}
  (a)~City reputation $R_c = \frac{1}{N_c}\sum_{i \in c} t_i\,(1 - b_i)$ per method (mean $\pm$ 95\%
  confidence interval).
  (b) to (e)~Each city's deviation from the mean reputation (diamond), decomposed exactly into a trust part,
  $(\bar{t}_c - \bar{t})(1 - \bar{b})$, a bias part, $-\bar{t}\,(\bar{b}_c - \bar{b})$, and a small
  remainder (interaction and covariance within cities).
  For the agent, trust accounts for about 90\% of the absolute city deviations, with Saint Louis,
  Philadelphia and New Orleans above and Saint Petersburg and Reno below the mean in every design. For the
  NLP, bias accounts for about 60\%, chiefly lifting New Orleans and Santa Barbara, the
  cities where it detects the least implicit bias.
  No city departs from the mean reputation by more than 0.03 on the scale from 0 to 1, a small
  margin compared with the spread of reputation across individual reviews.}
  \label{fig5}
\end{figure*}

The differences between cities are small for every method, but rankings of places derived from reviews are
published and used even when the underlying differences are small. We therefore ask not whether cities differ
substantially, but whether the two approaches agree on how reputable each city is and on why one city ranks
above another. Combining both measures into city reputation $R_c$ (Equation~\ref{eq:reputation}) shows that
they already differ in level (Figure~\ref{fig5}a). The agent places all cities between about 0.55 and 0.67,
depending on the design, whereas the NLP based method places them between about 0.38 and 0.41, because it
combines lower plausibility with higher implicit bias. This gap between methods is far larger than any
difference between cities within a method.

To identify what drives the differences between cities, we decompose each city's deviation from the mean
reputation of all reviews into a trust part, a bias part and a small remainder (Equation~\ref{eq:decomposition};
Figure~\ref{fig5}b to e). The two approaches attribute reputation to different mechanisms. For the agent, trust
accounts for about 90\% of the absolute city deviations. Saint Louis, Philadelphia and New Orleans lie above
the mean, and Saint Petersburg and Reno below it, in all three designs. For the NLP based method, bias accounts
for about 60\%. It lifts New Orleans and Santa Barbara above the mean chiefly because it detects less implicit
bias in their reviews. No city departs from the mean by more than 0.03 on the scale from 0 to 1. As a
consequence, the two approaches produce different reputational geographies. The agent designs rank the cities
by reputation almost identically (Spearman $\rho \geq 0.95$), whereas the NLP based method and the agent agree
only weakly (mean $\rho = 0.24$). A city that appears reputable under one instrument does so because of
characteristics the other instrument does not detect.

\pagebreak

\subsection{Reputation varies between reviews more than between cities}
\label{Reputation varies between reviews more than between cities}
 
\begin{figure*}[t]
  \centering
  \includegraphics[width=\textwidth]{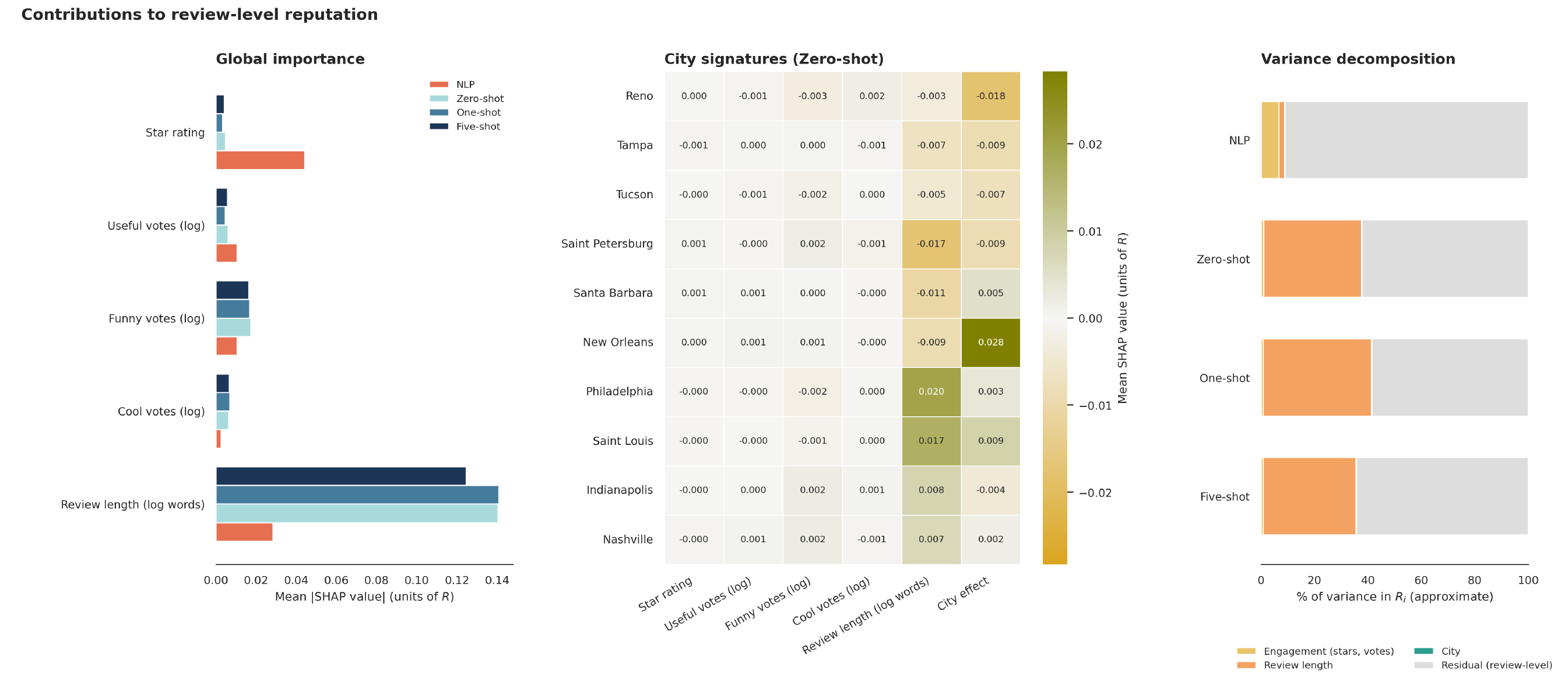}
  \caption{\textbf{Review length, not the city, explains the agent's reputation scores.}
  Contributions to review level reputation $R_i$ from a linear mixed model with standardized review
  characteristics and a random intercept per city.
  (a)~Global importance: mean absolute SHAP value, $\beta\,(x - \bar{x})$, per predictor. Review length
  dominates for all three agent designs (about 0.12-0.14 units of $R$), whereas star rating matters most
  for the NLP.
  (b)~City signatures for the zero-shot agent: mean SHAP value per city and predictor, together with the
  estimated city effect; cities are clustered by similarity.
  (c)~Approximate decomposition of the variance in $R_i$. Review length accounts for roughly 35-40\% of
  the agent's reputation variance, whereas cities account for 0.1-0.4\% (intra-class correlation,
  $\mathrm{ICC} = 0.001$-$0.004$ across methods).
  Reputation therefore varies far more between individual reviews than between the selected cities.}
  \label{fig6}
\end{figure*}

A multilevel model of review level reputation $R_i$, with standardized review characteristics and a random
intercept per city, strengthens the previous findings on how little the city itself contributes (Figure~\ref{fig6}). Between 0.1\% and
0.4\% of the variance in $R_i$ lies between cities (intra-class correlation, $\mathrm{ICC} =
0.001$-$0.004$ across methods). Five review characteristics, which are not themselves bias or trust scores, explain far more of the agent
based reputation scores (marginal $R^2 = 0.34$--$0.40$) than of the NLP based ones ($R^2 = 0.11$). These are
the star rating, the numbers of useful, funny and cool votes the review received, and its length in words,
with the vote counts and the length entered as logarithms. The city effects, that is, the estimated shift of
each city's reviews above or below the overall mean once these characteristics are taken into account, add
almost nothing, as the conditional $R^2$, which also counts them, is nearly equal to the marginal $R^2$. For
the zero-shot agent, they range from $-0.018$ for Reno to $+0.028$ for New Orleans on the scale from 0 to 1
(Figure~\ref{fig6}b), with Saint Louis ($+0.009$) and Santa Barbara ($+0.005$) also above and Tampa and Saint
Petersburg ($-0.009$) below the mean. None of the four city characteristics we examined, namely population, area, number of restaurants and
restaurant density, is significantly related to the city effects after Holm correction for any method. The largest associations, with restaurant density, range from $\rho = 0.31$ to $0.50$ across
methods, but with ten cities they cannot be distinguished from chance (Supplementary Material F).
 
Out of the selected review characteristics one characteristic dominates. Review length has the largest SHAP contribution for all three agent designs, about 0.12-0.14 units of $R$ on average, and accounts for roughly 35-40\% of the variance in the agent's reputation scores (Figure~\ref{fig6}a, c). One potential explanation for this is that longer reviews contain more of the concrete details the agent counts as evidence of credibility. For the NLP based method, star rating matters most, in line with its reliance on the sentiment of generic praise. The city signatures show the same picture at the level of cities (Figure~\ref{fig6}b). Because longer reviews
receive higher reputation scores from the agent, cities whose reviews are longer than average gain in
predicted reputation, and cities whose reviews are shorter lose. For the zero-shot agent, review length raises
the predicted reputation of Philadelphia by about 0.02 and of Saint Louis by about 0.017, and lowers that of
Saint Petersburg by about 0.017 and of Santa Barbara by about 0.011. These contributions are of the same size
as the largest city effects, which range from $-0.018$ to $+0.028$, and in Philadelphia and Saint Petersburg
they exceed the city effect itself.
 
These results answer the second research question. Each instrument derives a reputation for every city from
the reviews, but the two instruments derive different ones, in level, in ranking and in the mechanism to which
they attribute the differences. Within each instrument, the selected cities barely differ in reputation.
Variation lies mainly between individual reviews and is related to characteristics of the reviews such as
their length, which is to be expected, since both instruments score the text of the reviews. For platforms and
city marketing alike, a reputation score derived from reviews therefore depends on the instrument that
produces it, and cannot be read as a property of the city alone.
 
\clearpage

\section{Discussion}
\label{Discussion}
Agentic and corpus-based instruments give similar results for credibility but different results for bias and plausibility.
Cities, moreover, differ little in reputation within the same instrument, but the same city's reputation
differs across instruments. Research on destination image typically treats the image of a place, that is, how
visitors perceive and evaluate it, for example its quality, safety or authenticity, as a characteristic of the
place that reviews reveal \citep{marineroig2020, guo2021}. On this view, a better instrument would only measure
the same image more precisely. Our results question this assumption. In our data, the instrument shapes both
which cities appear reputable and why.

\subsection{What each instrument measures}
\label{What each instrument measures}

The two approaches converge where the construct is anchored in observable text. Credibility, defined by both as concrete, first-hand detail, yields substantial agreement ($\rho \approx 0.66$, $\kappa \approx 0.63$;
Figure~\ref{fig2}). Where the construct requires interpretation, they part ways. For plausibility, the two instruments measure different constructs under one name. The NLP-based model scores only logical consistency between pairs of sentences and finds a strong contradiction in a third of
the reviews, whereas the agents apply a broader definition that also covers realism and find almost no
problems. Averaging this dimension with credibility into one trust score hides the strong agreement on credibility behind the absent agreement on plausibility (Figure~\ref{fig2}b). Studies that report a single credibility or trust index from reviews risk
the same masking. Consequently, reporting the dimensions separately avoids it.
For bias, the corpus-based method is sensitive to generalization markers and complaint vocabulary (Figure~\ref{fig3}a), cues that also
appear in ordinary dissatisfaction. The agents instead quote references to social groups and places, the
vocabulary of the "discursive redlining" \citet{zukin2017} documented in Yelp reviews. Examined jointly, the NLP based method measures how strongly reviewers generalize and complain, whereas the agents measure whether they make claims about who people are. Neither result alone shows which is the valid measure of bias as the coding scheme defines it
\citep{jacobs2021}. The coding scheme defines bias as language that unfairly favors or disfavors individuals,
groups or entities, such as stereotypes about staff, customers, neighborhoods or the city, or sweeping claims
drawn from a single visit, and states that a strong opinion tied to a described experience is not bias
(Section~\ref{Agentic measurement}; Supplementary Material C). This follows research that locates bias in
language in unwarranted generalizations about social groups \citep{maass1999} and in framing that presents a
subjective evaluation as fact \citep{recasens2013}. The agents' quotes, which refer to social groups and
places, match the stereotypes named in this definition, whereas complaints such as \textit{rude} or
\textit{worst} fall under its exception for opinions tied to an experience. Generalization markers such as
\textit{never} and \textit{always}, however, can also signal the sweeping claims the definition counts as bias.
The contrast in the language each approach responds to therefore suggests that the agent's reading is closer to
the core of the definition, but establishing this requires human coding of the divergent reviews.

The language behind trust reveals a sharper conflict. The agents ground trust in concrete details and quotes
generic praise as a red flag, consistent with research on credibility cues \citep{filieri2016} and with
evidence that fabricated reviews contain fewer concrete details than genuine ones \citep{ott2011}. The corpus-based NLP, which considers details, first-person narration and length without judging their content, associates the same praise with high trust. An instrument that rewards
enthusiasm could favor exactly those reviews platforms try to filter out.

\subsection{Prompt design is part of the measurement}
\label{Prompt design is part of the measurement}

The agent designs agree closely with one another, so the divergence from the NLP is not an artefact of any single prompt. The examples nonetheless change what the agent counts as bias. With five examples, it
quotes superlatives such as \textit{best} and \textit{ever} as biased and estimates 21\% of reviews as biased instead of 7-8\% estimated by the zero and one-shot models
(Figure~\ref{fig3}a). The reason of this can be that one of the five examples used in the five-shot model (which has not been used in the one-shot model) scores an unsupported superlative (\textit{"the best pizza in the city,
nothing else even comes close"}) as mild bias (Supplementary Material C). The five-shot agent appears to have generalized this single case to ordinary
praise, a prompt design effect that any few-shot measurement of bias should test for. This is consistent with the known
sensitivity of language models to few-shot examples \citep{zhao2021, lu2022, sclar2024}, and it implies that 
examples must be validated as carefully as the coding scheme itself.

\subsection{Why cities barely differ}
\label{Why cities barely differ}

Less than half a percent of the variation in review level reputation lies between cities
(Figure~\ref{fig6}). Two mechanisms can explain this. First, reviews describe experiences with individual businesses, and these
vary far more within a city than between cities. Reviewers, moreover, evaluate these experiences relative to
what they expected rather than on an absolute scale. Their expectations are formed locally, by the price they
paid \citep{abrate2021} and by the earlier ratings of the same business, which act as an anchor for their own
\citep{wang2022anchoring}, so each review measures a business against local standards. Evaluations anchored in
this way are likely to vary around a similar average in every city, which leaves little room for differences
between cities. Second, reviewing is a genre with
shared conventions. People describe food, service and prices are in much the same way in Tucson as in Philadelphia. What varies relatively the most is how much a reviewer writes, and review length dominates the agents' reputation scores because longer reviews contain more of the detail the agents read as credibility.

The scale of the analysis matters here. The prejudiced framing documented by \citet{zukin2017} and the consumption segregation documented by \citet{davis2019}, \citet{moro2021} and \citet{hilman2022} operate between
neighborhoods, and they may cancel out in city averages. Our null result for cities therefore does not imply that reviews are free of place-based bias. Rather, it implies that city averages are too coarse to show it.

\subsection{Implications for platforms and cities}
\label{Implications for platforms and cities}

For a study that infers the image of places from reviews, our results call for treating the instrument as
part of the object of study. The same reviews yield different city rankings and attribute reputation to
different mechanisms, trust in case of the agent and bias in case of the NLP-based method (Figure~\ref{fig5}). Reporting both, with the decomposition into trust and bias, makes these choices visible. For platforms and city marketers, any reputation score computed from reviews inherits the values built into its instrument
\citep{orlikowski2014, gillespie2014}. If such scores are used to rank places or to moderate content, the instrument should be documented and audited like any other part of the platform's valuation apparatus.

\subsection{Limitations}
\label{Limitations}

Several limitations bound these conclusions. The study covers one
platform, ten US\ cities and English-language reviews. The agents are run on a single open-weight model once per review at low temperature; other models may draw the construct boundaries differently. Furthermore, language models carry social biases of their own that can shape how they judge bias \citep{gallegos2024}. The examples used in the prompts are synthetic, and at least one of them shifted the five-shot agent's notion of bias. The corpus-based cues uses thresholds fixed apriori, while other thresholds would change its levels and possibly its agreement with the agent. Finally, without human annotation of the same reviews, we can show where the instruments agree and why they diverge,
but not which one is closer to human judgment.

Each limitation suggests a next step. Human coding of a stratified sample of convergent and divergent reviews would give opportunity for the measurement of validity for both instruments. Random or business-stratified sampling, analyses at the neighborhood scale, and replication across models, platforms and other cities would test how far the near-absence of city differences generalizes. The examples drawn from real, human-coded reviews, and a test of how each example
shifts the agent's judgments, would make few-shot measurement auditable.

\section{Conclusion}
We compared an agentic and a corpus-based instrument for measuring trust and bias in  100{,}000 Yelp
reviews from ten US\ cities. The two agree on credibility, the dimension most firmly anchored in the text, and diverge on bias and plausibility, where judgment requires interpretation. They respond to different language features: the corpus-based NLP to complaints and generalizations, and the agent to statements about social
groups. Aggregated to cities, both types of models found little variation, and the small differences that remain arise from trust under one instrument and from bias under the other.

In the ten cities studied, online reviews, thus yield little difference in reputation between cities. It is shown that how a reputation emerges from them depends on the angle from which the reviews are read. As platforms and city marketers increasingly delegate
that reading to language models, the definitions, examples and checks built into these models become part of how places are valued. Making those choices explicit, and testing them against human judgment and at the
scale of neighborhoods, is the next step towards reputation measures that could reveal the real characteristics of places rather than the
instruments that score them.\\

\textbf{CRediT authorship contribution statement}\\
\textbf{Rafiazka Hilman}: Conceptualization, Methodology, Software, Validation, Formal analysis, Investigation, Resources, Data Curation, Writing – Original
Draft, Writing – Review \& Editing,  Visualization, Funding acquisition\\ 
\textbf{Júlia Koltai}: Conceptualization, Validation, Resources, Writing – Review \& Editing, Supervision, Project administration, Funding acquisition\\

\textbf{Declaration of competing interest}\\
The authors declare that they have no known competing financial interests or personal relationships that could have appeared to influence the work reported in this paper.\\

\textbf{Funding sources}\\
Rafiazka Hilman acknowledges support from the Momentum MSCA Programme co-funded by the European Commission through the HORIZON-MSCA-2023-COFUND programme and the Secretariat of the Hungarian Academy of Sciences (MTA) under Grant Agreement No. 101179854. Julia Koltai acknowledges funding from the
Hungarian Academy of Sciences Lendület Program: LP2022-10/2022. \\

\clearpage

\newpage

\bibliography{sample}

@article{babicrosario2020,
  author  = {Babi{\'c} Rosario, Ana and de Valck, Kristine and Sotgiu, Francesca},
  title   = {Conceptualizing the Electronic Word-of-Mouth Process: What We Know and Need to Know about {eWOM} Creation, Exposure, and Evaluation},
  journal = {Journal of the Academy of Marketing Science},
  year    = {2020},
  volume  = {48},
  number  = {3},
  pages   = {422--448},
  doi     = {10.1007/s11747-019-00706-1}
}

@article{bae2011,
  author  = {Bae, Soonyong and Lee, Taesik},
  title   = {Product Type and Consumers' Perception of Online Consumer Reviews},
  journal = {Electronic Markets},
  year    = {2011},
  volume  = {21},
  number  = {4},
  pages   = {255--266},
  doi     = {10.1007/s12525-011-0072-0}
}

@techreport{luca2016,
  author      = {Luca, Michael},
  title       = {Reviews, Reputation, and Revenue: The Case of {Yelp.com}},
  institution = {Harvard Business School},
  type        = {Working Paper},
  number      = {12-016},
  year        = {2016}
}

@article{mayzlin2014,
  author  = {Mayzlin, Dina and Dover, Yaniv and Chevalier, Judith},
  title   = {Promotional Reviews: An Empirical Investigation of Online Review Manipulation},
  journal = {American Economic Review},
  year    = {2014},
  volume  = {104},
  number  = {8},
  pages   = {2421--2455},
  doi     = {10.1257/aer.104.8.2421}
}

@article{he2022,
  author  = {He, Sherry and Hollenbeck, Brett and Proserpio, Davide},
  title   = {The Market for Fake Reviews},
  journal = {Marketing Science},
  year    = {2022},
  volume  = {41},
  number  = {5},
  pages   = {896--921}
}

@article{filieri2016,
  author  = {Filieri, Raffaele},
  title   = {What Makes an Online Consumer Review Trustworthy?},
  journal = {Annals of Tourism Research},
  year    = {2016},
  volume  = {58},
  pages   = {46--64},
  doi     = {10.1016/j.annals.2015.12.019}
}

@article{zelenka2021,
  author  = {Zelenka, Josef and Azubuike, Tracy and P{\'a}skov{\'a}, Martina},
  title   = {Trust Model for Online Reviews of Tourism Services and Evaluation of Destinations},
  journal = {Administrative Sciences},
  year    = {2021},
  volume  = {11},
  number  = {2},
  pages   = {34},
  doi     = {10.3390/admsci11020034}
}

@article{marineroig2020,
  author  = {Marine-Roig, Estela and Huertas, Assumpci{\'o}},
  title   = {How Safety Affects Destination Image Projected through Online Travel Reviews},
  journal = {Journal of Destination Marketing \& Management},
  year    = {2020},
  volume  = {18},
  pages   = {100469},
  doi     = {10.1016/j.jdmm.2020.100469}
}

@article{guo2021,
  author  = {Guo, Xinxin and Pesonen, Juho and Komppula, Raija},
  title   = {Comparing Online Travel Review Platforms as Destination Image Information Agents},
  journal = {Information Technology \& Tourism},
  year    = {2021},
  volume  = {23},
  number  = {2},
  pages   = {159--187},
  doi     = {10.1007/s40558-021-00201-w}
}

@article{zukin2017,
  author  = {Zukin, Sharon and Lindeman, Scarlett and Hurson, Laurie},
  title   = {The Omnivore's Neighborhood? {Online} Restaurant Reviews, Race, and Gentrification},
  journal = {Journal of Consumer Culture},
  year    = {2017},
  volume  = {17},
  number  = {3},
  pages   = {459--479},
  doi     = {10.1177/1469540515611203}
}

@inproceedings{recasens2013,
  author    = {Recasens, Marta and Danescu-Niculescu-Mizil, Cristian and Jurafsky, Dan},
  title     = {Linguistic Models for Analyzing and Detecting Biased Language},
  booktitle = {Proceedings of the 51st Annual Meeting of the Association for Computational Linguistics},
  year      = {2013},
  pages     = {1650--1659}
}

@article{caliskan2017,
  author  = {Caliskan, Aylin and Bryson, Joanna J. and Narayanan, Arvind},
  title   = {Semantics Derived Automatically from Language Corpora Contain Human-like Biases},
  journal = {Science},
  year    = {2017},
  volume  = {356},
  number  = {6334},
  pages   = {183--186},
  doi     = {10.1126/science.aal4230}
}

@inproceedings{blodgett2020,
  author    = {Blodgett, Su Lin and Barocas, Solon and Daum{\'e} III, Hal and Wallach, Hanna},
  title     = {Language (Technology) is Power: A Critical Survey of ``Bias'' in {NLP}},
  booktitle = {Proceedings of the 58th Annual Meeting of the Association for Computational Linguistics},
  year      = {2020},
  pages     = {5454--5476}
}

@article{monroe2008,
  author  = {Monroe, Burt L. and Colaresi, Michael P. and Quinn, Kevin M.},
  title   = {Fightin' Words: Lexical Feature Selection and Evaluation for Identifying the Content of Political Conflict},
  journal = {Political Analysis},
  year    = {2008},
  volume  = {16},
  number  = {4},
  pages   = {372--403},
  doi     = {10.1093/pan/mpn018}
}

@article{gilardi2023,
  author  = {Gilardi, Fabrizio and Alizadeh, Meysam and Kubli, Ma{\"e}l},
  title   = {{ChatGPT} Outperforms Crowd Workers for Text-Annotation Tasks},
  journal = {Proceedings of the National Academy of Sciences},
  year    = {2023},
  volume  = {120},
  number  = {30},
  pages   = {e2305016120},
  doi     = {10.1073/pnas.2305016120}
}

@article{ziems2024,
  author  = {Ziems, Caleb and Held, William and Shaikh, Omar and Chen, Jiaao and Zhang, Zhehao and Yang, Diyi},
  title   = {Can Large Language Models Transform Computational Social Science?},
  journal = {Computational Linguistics},
  year    = {2024},
  volume  = {50},
  number  = {1},
  pages   = {237--291},
  doi     = {10.1162/coli_a_00502}
}

@inproceedings{madaan2023,
  author    = {Madaan, Aman and Tandon, Niket and Gupta, Prakhar and Hallinan, Skyler and Gao, Luyu and Wiegreffe, Sarah and Alon, Uri and Dziri, Nouha and Prabhumoye, Shrimai and Yang, Yiming and Gupta, Shashank and Majumder, Bodhisattwa Prasad and Hermann, Katherine and Welleck, Sean and Yazdanbakhsh, Amir and Clark, Peter},
  title     = {Self-Refine: Iterative Refinement with Self-Feedback},
  booktitle = {Advances in Neural Information Processing Systems},
  year      = {2023},
  volume    = {36}
}

@article{wang2024,
  author  = {Wang, Lei and Ma, Chen and Feng, Xueyang and Zhang, Zeyu and Yang, Hao and Zhang, Jingsen and Chen, Zhiyuan and Tang, Jiakai and Chen, Xu and Lin, Yankai and Zhao, Wayne Xin and Wei, Zhewei and Wen, Jirong},
  title   = {A Survey on Large Language Model Based Autonomous Agents},
  journal = {Frontiers of Computer Science},
  year    = {2024},
  volume  = {18},
  number  = {6},
  pages   = {186345},
  doi     = {10.1007/s11704-024-40231-1}
}

@misc{yelpdataset,
  author       = {{Yelp Inc.}},
  title        = {Yelp Open Dataset},
  howpublished = {\url{https://www.yelp.com/dataset}},
  year         = {2026},
  note         = {Accessed [DATE]}
}

@inproceedings{sadvilkar2020,
  author    = {Sadvilkar, Nipun and Neumann, Mark},
  title     = {{PySBD}: Pragmatic Sentence Boundary Disambiguation},
  booktitle = {Proceedings of the Second Workshop for NLP Open Source Software (NLP-OSS)},
  year      = {2020},
  pages     = {110--114}
}

@misc{honnibal2020,
  author       = {Honnibal, Matthew and Montani, Ines and Van Landeghem, Sofie and Boyd, Adriane},
  title        = {{spaCy}: Industrial-Strength Natural Language Processing in {Python}},
  year         = {2020},
  howpublished = {Zenodo},
  doi          = {10.5281/zenodo.1212303}
}

@misc{detoxify2020,
  author       = {Hanu, Laura and {Unitary team}},
  title        = {Detoxify},
  year         = {2020},
  howpublished = {\url{https://github.com/unitaryai/detoxify}}
}

@inproceedings{reimers2019,
  author    = {Reimers, Nils and Gurevych, Iryna},
  title     = {Sentence-{BERT}: Sentence Embeddings using {Siamese} {BERT}-Networks},
  booktitle = {Proceedings of the 2019 Conference on Empirical Methods in Natural Language Processing and the 9th International Joint Conference on Natural Language Processing},
  year      = {2019},
  pages     = {3982--3992}
}

@inproceedings{loureiro2022,
  author    = {Loureiro, Daniel and Barbieri, Francesco and Neves, Leonardo and Espinosa Anke, Luis and Camacho-Collados, Jose},
  title     = {{TimeLMs}: Diachronic Language Models from {Twitter}},
  booktitle = {Proceedings of the 60th Annual Meeting of the Association for Computational Linguistics: System Demonstrations},
  year      = {2022},
  pages     = {251--260}
}

@article{laurer2024,
  author  = {Laurer, Moritz and van Atteveldt, Wouter and Casas, Andreu and Welbers, Kasper},
  title   = {Less Annotating, More Classifying: Addressing the Data Scarcity Issue of Supervised Machine Learning with Deep Transfer Learning and {BERT-NLI}},
  journal = {Political Analysis},
  year    = {2024},
  volume  = {32},
  number  = {1},
  pages   = {84--100},
  doi     = {10.1017/pan.2023.20}
}

@article{cohen1968,
  author  = {Cohen, Jacob},
  title   = {Weighted Kappa: Nominal Scale Agreement with Provision for Scaled Disagreement or Partial Credit},
  journal = {Psychological Bulletin},
  year    = {1968},
  volume  = {70},
  number  = {4},
  pages   = {213--220},
  doi     = {10.1037/h0026256}
}

@article{landis1977,
  author  = {Landis, J. Richard and Koch, Gary G.},
  title   = {The Measurement of Observer Agreement for Categorical Data},
  journal = {Biometrics},
  year    = {1977},
  volume  = {33},
  number  = {1},
  pages   = {159--174},
  doi     = {10.2307/2529310}
}

@article{vargha2000,
  author  = {Vargha, Andr{\'a}s and Delaney, Harold D.},
  title   = {A Critique and Improvement of the {CL} Common Language Effect Size Statistics of {McGraw} and {Wong}},
  journal = {Journal of Educational and Behavioral Statistics},
  year    = {2000},
  volume  = {25},
  number  = {2},
  pages   = {101--132},
  doi     = {10.3102/10769986025002101}
}

@article{holm1979,
  author  = {Holm, Sture},
  title   = {A Simple Sequentially Rejective Multiple Test Procedure},
  journal = {Scandinavian Journal of Statistics},
  year    = {1979},
  volume  = {6},
  number  = {2},
  pages   = {65--70}
}

@article{tomczak2014,
  author  = {Tomczak, Maciej and Tomczak, Ewa},
  title   = {The Need to Report Effect Size Estimates Revisited: An Overview of Some Recommended Measures of Effect Size},
  journal = {Trends in Sport Sciences},
  year    = {2014},
  volume  = {21},
  number  = {1},
  pages   = {19--25}
}

@article{nakagawa2013,
  author  = {Nakagawa, Shinichi and Schielzeth, Holger},
  title   = {A General and Simple Method for Obtaining {$R^2$} from Generalized Linear Mixed-Effects Models},
  journal = {Methods in Ecology and Evolution},
  year    = {2013},
  volume  = {4},
  number  = {2},
  pages   = {133--142},
  doi     = {10.1111/j.2041-210x.2012.00261.x}
}

@inproceedings{lundberg2017,
  author    = {Lundberg, Scott M. and Lee, Su-In},
  title     = {A Unified Approach to Interpreting Model Predictions},
  booktitle = {Advances in Neural Information Processing Systems},
  year      = {2017},
  volume    = {30}
}

@article{resnick2000,
  author  = {Resnick, Paul and Kuwabara, Ko and Zeckhauser, Richard and Friedman, Eric},
  title   = {Reputation Systems},
  journal = {Communications of the ACM},
  year    = {2000},
  volume  = {43},
  number  = {12},
  pages   = {45--48},
  doi     = {10.1145/355112.355122}
}

@article{chevalier2006,
  author  = {Chevalier, Judith A. and Mayzlin, Dina},
  title   = {The Effect of Word of Mouth on Sales: Online Book Reviews},
  journal = {Journal of Marketing Research},
  year    = {2006},
  volume  = {43},
  number  = {3},
  pages   = {345--354},
  doi     = {10.1509/jmkr.43.3.345}
}

@article{orlikowski2014,
  author  = {Orlikowski, Wanda J. and Scott, Susan V.},
  title   = {What Happens When Evaluation Goes Online? {Exploring} Apparatuses of Valuation in the Travel Sector},
  journal = {Organization Science},
  year    = {2014},
  volume  = {25},
  number  = {3},
  pages   = {868--891},
  doi     = {10.1287/orsc.2013.0877}
}

@book{vandijck2018,
  author    = {van Dijck, Jos{\'e} and Poell, Thomas and de Waal, Martijn},
  title     = {The Platform Society: Public Values in a Connective World},
  publisher = {Oxford University Press},
  address   = {New York},
  year      = {2018}
}

@incollection{gillespie2014,
  author    = {Gillespie, Tarleton},
  title     = {The Relevance of Algorithms},
  booktitle = {Media Technologies: Essays on Communication, Materiality, and Society},
  editor    = {Gillespie, Tarleton and Boczkowski, Pablo J. and Foot, Kirsten A.},
  publisher = {MIT Press},
  address   = {Cambridge, MA},
  year      = {2014},
  pages     = {167--194}
}

@article{kavaratzis2013,
  author  = {Kavaratzis, Mihalis and Hatch, Mary Jo},
  title   = {The Dynamics of Place Brands: An Identity-Based Approach to Place Branding Theory},
  journal = {Marketing Theory},
  year    = {2013},
  volume  = {13},
  number  = {1},
  pages   = {69--86},
  doi     = {10.1177/1470593112467268}
}

@article{glaeser2018,
  author  = {Glaeser, Edward L. and Kim, Hyunjin and Luca, Michael},
  title   = {Nowcasting Gentrification: Using {Yelp} Data to Quantify Neighborhood Change},
  journal = {AEA Papers and Proceedings},
  year    = {2018},
  volume  = {108},
  pages   = {77--82},
  doi     = {10.1257/pandp.20181034}
}

@article{davis2019,
  author  = {Davis, Donald R. and Dingel, Jonathan I. and Monras, Joan and Morales, Eduardo},
  title   = {How Segregated Is Urban Consumption?},
  journal = {Journal of Political Economy},
  year    = {2019},
  volume  = {127},
  number  = {4},
  pages   = {1684--1738},
  doi     = {10.1086/701680}
}

@article{hilman2022,
  author  = {Hilman, Rafiazka Millanida and I{\~n}iguez, Gerardo and Karsai, M{\'a}rton},
  title   = {Socioeconomic Biases in Urban Mixing Patterns of {US} Metropolitan Areas},
  journal = {EPJ Data Science},
  year    = {2022},
  volume  = {11},
  pages   = {32},
  doi     = {10.1140/epjds/s13688-022-00341-x}
}

@article{metzger2010,
  author  = {Metzger, Miriam J. and Flanagin, Andrew J. and Medders, Ryan B.},
  title   = {Social and Heuristic Approaches to Credibility Evaluation Online},
  journal = {Journal of Communication},
  year    = {2010},
  volume  = {60},
  number  = {3},
  pages   = {413--439},
  doi     = {10.1111/j.1460-2466.2010.01488.x}
}

@inproceedings{ott2011,
  author    = {Ott, Myle and Choi, Yejin and Cardie, Claire and Hancock, Jeffrey T.},
  title     = {Finding Deceptive Opinion Spam by Any Stretch of the Imagination},
  booktitle = {Proceedings of the 49th Annual Meeting of the Association for Computational Linguistics: Human Language Technologies},
  year      = {2011},
  pages     = {309--319}
}

@article{garg2018,
  author  = {Garg, Nikhil and Schiebinger, Londa and Jurafsky, Dan and Zou, James},
  title   = {Word Embeddings Quantify 100 Years of Gender and Ethnic Stereotypes},
  journal = {Proceedings of the National Academy of Sciences},
  year    = {2018},
  volume  = {115},
  number  = {16},
  pages   = {E3635--E3644},
  doi     = {10.1073/pnas.1720347115}
}

@inproceedings{davidson2019,
  author    = {Davidson, Thomas and Bhattacharya, Debasmita and Weber, Ingmar},
  title     = {Racial Bias in Hate Speech and Abusive Language Detection Datasets},
  booktitle = {Proceedings of the Third Workshop on Abusive Language Online},
  year      = {2019},
  pages     = {25--35}
}

@article{hovy2021,
  author  = {Hovy, Dirk and Prabhumoye, Shrimai},
  title   = {Five Sources of Bias in Natural Language Processing},
  journal = {Language and Linguistics Compass},
  year    = {2021},
  volume  = {15},
  number  = {8},
  pages   = {e12432},
  doi     = {10.1111/lnc3.12432}
}

@inproceedings{jacobs2021,
  author    = {Jacobs, Abigail Z. and Wallach, Hanna},
  title     = {Measurement and Fairness},
  booktitle = {Proceedings of the 2021 ACM Conference on Fairness, Accountability, and Transparency},
  year      = {2021},
  pages     = {375--385},
  doi       = {10.1145/3442188.3445901}
}

@book{grimmer2022,
  author    = {Grimmer, Justin and Roberts, Margaret E. and Stewart, Brandon M.},
  title     = {Text as Data: A New Framework for Machine Learning and the Social Sciences},
  publisher = {Princeton University Press},
  address   = {Princeton, NJ},
  year      = {2022}
}

@article{argyle2023,
  author  = {Argyle, Lisa P. and Busby, Ethan C. and Fulda, Nancy and Gubler, Joshua R. and Rytting, Christopher and Wingate, David},
  title   = {Out of One, Many: Using Language Models to Simulate Human Samples},
  journal = {Political Analysis},
  year    = {2023},
  volume  = {31},
  number  = {3},
  pages   = {337--351},
  doi     = {10.1017/pan.2023.2}
}

@article{bail2024,
  author  = {Bail, Christopher A.},
  title   = {Can Generative {AI} Improve Social Science?},
  journal = {Proceedings of the National Academy of Sciences},
  year    = {2024},
  volume  = {121},
  number  = {21},
  pages   = {e2314021121},
  doi     = {10.1073/pnas.2314021121}
}

@inproceedings{zhao2021,
  author    = {Zhao, Zihao and Wallace, Eric and Feng, Shi and Klein, Dan and Singh, Sameer},
  title     = {Calibrate Before Use: Improving Few-Shot Performance of Language Models},
  booktitle = {Proceedings of the 38th International Conference on Machine Learning},
  series    = {Proceedings of Machine Learning Research},
  volume    = {139},
  year      = {2021},
  pages     = {12697--12706}
}

@inproceedings{lu2022,
  author    = {Lu, Yao and Bartolo, Max and Moore, Alastair and Riedel, Sebastian and Stenetorp, Pontus},
  title     = {Fantastically Ordered Prompts and Where to Find Them: Overcoming Few-Shot Prompt Order Sensitivity},
  booktitle = {Proceedings of the 60th Annual Meeting of the Association for Computational Linguistics},
  year      = {2022},
  pages     = {8086--8098}
}

@inproceedings{zheng2023,
  author    = {Zheng, Lianmin and Chiang, Wei-Lin and Sheng, Ying and Zhuang, Siyuan and Wu, Zhanghao and Zhuang, Yonghao and Lin, Zi and Li, Zhuohan and Li, Dacheng and Xing, Eric P. and Zhang, Hao and Gonzalez, Joseph E. and Stoica, Ion},
  title     = {Judging {LLM}-as-a-Judge with {MT-Bench} and {Chatbot Arena}},
  booktitle = {Advances in Neural Information Processing Systems},
  volume    = {36},
  year      = {2023}
}

@inproceedings{huang2024,
  author    = {Huang, Jie and Chen, Xinyun and Mishra, Swaroop and Zheng, Huaixiu Steven and Yu, Adams Wei and Song, Xinying and Zhou, Denny},
  title     = {Large Language Models Cannot Self-Correct Reasoning Yet},
  booktitle = {International Conference on Learning Representations},
  year      = {2024}
}

@misc{gemma2026,
  author        = {{Gemma Team}},
  title         = {Gemma 4 Technical Report},
  year          = {2026},
  eprint        = {2607.02770},
  archivePrefix = {arXiv},
  primaryClass  = {cs.CL},
  doi           = {10.48550/arXiv.2607.02770}
}

@article{luca2016fraud,
  author  = {Luca, Michael and Zervas, Georgios},
  title   = {Fake It Till You Make It: Reputation, Competition, and {Yelp} Review Fraud},
  journal = {Management Science},
  year    = {2016},
  volume  = {62},
  number  = {12},
  pages   = {3412--3427},
  doi     = {10.1287/mnsc.2015.2304}
}

@article{babicrosario2016,
  author  = {Babi{\'c} Rosario, Ana and Sotgiu, Francesca and De Valck, Kristine and Bijmolt, Tammo H. A.},
  title   = {The Effect of Electronic Word of Mouth on Sales: A Meta-Analytic Review of Platform, Product, and Metric Factors},
  journal = {Journal of Marketing Research},
  year    = {2016},
  volume  = {53},
  number  = {3},
  pages   = {297--318},
  doi     = {10.1509/jmr.14.0380}
}

@article{rathje2024,
  author  = {Rathje, Steve and Mirea, Dan-Mircea and Sucholutsky, Ilia and Marjieh, Raja and Robertson, Claire E. and Van Bavel, Jay J.},
  title   = {{GPT} is an Effective Tool for Multilingual Psychological Text Analysis},
  journal = {Proceedings of the National Academy of Sciences},
  year    = {2024},
  volume  = {121},
  number  = {34},
  pages   = {e2308950121},
  doi     = {10.1073/pnas.2308950121}
}

@inproceedings{sclar2024,
  author    = {Sclar, Melanie and Choi, Yejin and Tsvetkov, Yulia and Suhr, Alane},
  title     = {Quantifying Language Models' Sensitivity to Spurious Features in Prompt Design or: How {I} Learned to Start Worrying about Prompt Formatting},
  booktitle = {International Conference on Learning Representations},
  year      = {2024}
}

@article{kamoi2024,
  author  = {Kamoi, Ryo and Zhang, Yusen and Zhang, Nan and Han, Jiawei and Zhang, Rui},
  title   = {When Can {LLMs} Actually Correct Their Own Mistakes? {A} Critical Survey of Self-Correction of {LLMs}},
  journal = {Transactions of the Association for Computational Linguistics},
  year    = {2024},
  volume  = {12},
  pages   = {1417--1440},
  doi     = {10.1162/tacl_a_00713}
}

@inproceedings{sap2019,
  author    = {Sap, Maarten and Card, Dallas and Gabriel, Saadia and Choi, Yejin and Smith, Noah A.},
  title     = {The Risk of Racial Bias in Hate Speech Detection},
  booktitle = {Proceedings of the 57th Annual Meeting of the Association for Computational Linguistics},
  year      = {2019},
  pages     = {1668--1678},
  doi       = {10.18653/v1/P19-1163}
}

@article{moro2021,
  author  = {Moro, Esteban and Calacci, Dan and Dong, Xiaowen and Pentland, Alex},
  title   = {Mobility Patterns Are Associated with Experienced Income Segregation in Large {US} Cities},
  journal = {Nature Communications},
  year    = {2021},
  volume  = {12},
  pages   = {4633},
  doi     = {10.1038/s41467-021-24899-8}
}

@article{gallegos2024,
  author  = {Gallegos, Isabel O. and Rossi, Ryan A. and Barrow, Joe and Tanjim, Md Mehrab and Kim, Sungchul and Dernoncourt, Franck and Yu, Tong and Zhang, Ruiyi and Ahmed, Nesreen K.},
  title   = {Bias and Fairness in Large Language Models: A Survey},
  journal = {Computational Linguistics},
  year    = {2024},
  volume  = {50},
  number  = {3},
  pages   = {1097--1179},
  doi     = {10.1162/coli_a_00524}
}

@article{ismagilova2020,
  author  = {Ismagilova, Elvira and Slade, Emma and Rana, Nripendra P. and Dwivedi, Yogesh K.},
  title   = {The Effect of Characteristics of Source Credibility on Consumer Behaviour: A Meta-Analysis},
  journal = {Journal of Retailing and Consumer Services},
  year    = {2020},
  volume  = {53},
  pages   = {101736}
}

@article{erkan2016,
  author  = {Erkan, Ismail and Evans, Chris},
  title   = {The Influence of {eWOM} in Social Media on Consumers' Purchase Intentions: An Extended Approach to Information Adoption},
  journal = {Computers in Human Behavior},
  year    = {2016},
  volume  = {61},
  pages   = {47--55},
  doi     = {10.1016/j.chb.2016.03.003}
}

@article{schoenmueller2020,
  author  = {Schoenmueller, Verena and Netzer, Oded and Stahl, Florian},
  title   = {The Polarity of Online Reviews: Prevalence, Drivers and Implications},
  journal = {Journal of Marketing Research},
  year    = {2020},
  volume  = {57},
  number  = {5},
  pages   = {853--877},
  doi     = {10.1177/0022243720941832}
}

@inproceedings{hutto2014,
  author    = {Hutto, C. J. and Gilbert, Eric},
  title     = {{VADER}: A Parsimonious Rule-Based Model for Sentiment Analysis of Social Media Text},
  booktitle = {Proceedings of the International AAAI Conference on Web and Social Media},
  year      = {2014},
  volume    = {8},
  pages     = {216--225}
}

@inproceedings{williams2018,
  author    = {Williams, Adina and Nangia, Nikita and Bowman, Samuel},
  title     = {A Broad-Coverage Challenge Corpus for Sentence Understanding through Inference},
  booktitle = {Proceedings of the 2018 Conference of the North American Chapter of the Association for Computational Linguistics: Human Language Technologies},
  year      = {2018},
  pages     = {1112--1122},
  doi       = {10.18653/v1/N18-1101}
}

@inproceedings{deyoung2020,
  author    = {DeYoung, Jay and Jain, Sarthak and Rajani, Nazneen Fatema and Lehman, Eric and Xiong, Caiming and Socher, Richard and Wallace, Byron C.},
  title     = {{ERASER}: A Benchmark to Evaluate Rationalized {NLP} Models},
  booktitle = {Proceedings of the 58th Annual Meeting of the Association for Computational Linguistics},
  year      = {2020},
  pages     = {4443--4458},
  doi       = {10.18653/v1/2020.acl-main.408}
}

@inproceedings{gao2023,
  author    = {Gao, Tianyu and Yen, Howard and Yu, Jiatong and Chen, Danqi},
  title     = {Enabling Large Language Models to Generate Text with Citations},
  booktitle = {Proceedings of the 2023 Conference on Empirical Methods in Natural Language Processing},
  year      = {2023},
  pages     = {6465--6488},
  doi       = {10.18653/v1/2023.emnlp-main.398}
}

@article{hong2017,
  author  = {Hong, Hong and Xu, Di and Wang, G. Alan and Fan, Weiguo},
  title   = {Understanding the Determinants of Online Review Helpfulness: A Meta-Analytic Investigation},
  journal = {Decision Support Systems},
  year    = {2017},
  volume  = {102},
  pages   = {1--11},
  doi     = {10.1016/j.dss.2017.06.007}
}

@article{lo2019,
  author  = {Lo, Ada S. and Yao, Sharon Siyu},
  title   = {What Makes Hotel Online Reviews Credible? An Investigation of the Roles of Reviewer Expertise, Review Rating Consistency and Review Valence},
  journal = {International Journal of Contemporary Hospitality Management},
  year    = {2019},
  volume  = {31},
  number  = {1},
  pages   = {41--60},
  doi     = {10.1108/IJCHM-10-2017-0671}
}

@article{chen2013,
  author  = {Chen, Zoey and Lurie, Nicholas H.},
  title   = {Temporal Contiguity and Negativity Bias in the Impact of Online Word of Mouth},
  journal = {Journal of Marketing Research},
  year    = {2013},
  volume  = {50},
  number  = {4},
  pages   = {463--476}
}

@article{semin1988,
  author  = {Semin, G{\"u}n R. and Fiedler, Klaus},
  title   = {The Cognitive Functions of Linguistic Categories in Describing Persons: Social Cognition and Language},
  journal = {Journal of Personality and Social Psychology},
  year    = {1988},
  volume  = {54},
  number  = {4},
  pages   = {558--568}
}

@incollection{maass1999,
  author    = {Maass, Anne},
  title     = {Linguistic Intergroup Bias: Stereotype Perpetuation through Language},
  booktitle = {Advances in Experimental Social Psychology},
  editor    = {Zanna, Mark P.},
  volume    = {31},
  publisher = {Academic Press},
  year      = {1999},
  pages     = {79--121}
}

@inproceedings{borkan2019,
  author    = {Borkan, Daniel and Dixon, Lucas and Sorensen, Jeffrey and Thain, Nithum and Vasserman, Lucy},
  title     = {Nuanced Metrics for Measuring Unintended Bias with Real Data for Text Classification},
  booktitle = {Companion Proceedings of the 2019 World Wide Web Conference},
  year      = {2019},
  pages     = {491--500}
}

@article{kitagawa1955,
  author  = {Kitagawa, Evelyn M.},
  title   = {Components of a Difference Between Two Rates},
  journal = {Journal of the American Statistical Association},
  year    = {1955},
  volume  = {50},
  number  = {272},
  pages   = {1168--1194}
}

@book{dasgupta1993,
  author    = {Das Gupta, Prithwis},
  title     = {Standardization and Decomposition of Rates: A User's Manual},
  series    = {Current Population Reports, Series P23-186},
  publisher = {U.S. Bureau of the Census},
  address   = {Washington, DC},
  year      = {1993}
}

@article{elff2021,
  author  = {Elff, Martin and Heisig, Jan Paul and Schaeffer, Merlin and Shikano, Susumu},
  title   = {Multilevel Analysis with Few Clusters: Improving Likelihood-Based Methods to Provide Unbiased Estimates and Accurate Inference},
  journal = {British Journal of Political Science},
  year    = {2021},
  volume  = {51},
  number  = {1},
  pages   = {412--426}
}

@article{maas2005,
  author  = {Maas, Cora J. M. and Hox, Joop J.},
  title   = {Sufficient Sample Sizes for Multilevel Modeling},
  journal = {Methodology},
  year    = {2005},
  volume  = {1},
  number  = {3},
  pages   = {86--92},
  doi     = {10.1027/1614-2241.1.3.86}
}

@article{mcneish2016,
  author  = {McNeish, Daniel M. and Stapleton, Laura M.},
  title   = {Modeling Clustered Data with Very Few Clusters},
  journal = {Multivariate Behavioral Research},
  year    = {2016},
  volume  = {51},
  number  = {4},
  pages   = {495--518}
}

@article{bryan2016,
  author  = {Bryan, Mark L. and Jenkins, Stephen P.},
  title   = {Multilevel Modelling of Country Effects: A Cautionary Tale},
  journal = {European Sociological Review},
  year    = {2016},
  volume  = {32},
  number  = {1},
  pages   = {3--22}
}

@article{cameron2008,
  author  = {Cameron, A. Colin and Gelbach, Jonah B. and Miller, Douglas L.},
  title   = {Bootstrap-Based Improvements for Inference with Clustered Errors},
  journal = {The Review of Economics and Statistics},
  year    = {2008},
  volume  = {90},
  number  = {3},
  pages   = {414--427}
}

@book{gelman2020,
  author    = {Gelman, Andrew and Hill, Jennifer and Vehtari, Aki},
  title     = {Regression and Other Stories},
  publisher = {Cambridge University Press},
  address   = {Cambridge},
  year      = {2020}
}

@article{abrate2021,
  author  = {Abrate, Graziano and Quinton, Sarah and Pera, Rebecca},
  title   = {The Relationship between Price Paid and Hotel Review Ratings: Expectancy-Disconfirmation or Placebo Effect?},
  journal = {Tourism Management},
  year    = {2021},
  volume  = {85},
  pages   = {104314},
  doi     = {10.1016/j.tourman.2021.104314}
}

@article{wang2022anchoring,
  author  = {Wang, Qian and Chau, Michael and Peng, Chih-Hung and Ngai, Eric W. T.},
  title   = {Using the Anchoring Effect and the Cultural Dimensions Theory to Study Customers' Online Rating Behaviors},
  journal = {Information Systems Frontiers},
  year    = {2022},
  volume  = {24},
  number  = {5},
  pages   = {1451--1463},
  doi     = {10.1007/s10796-021-10148-2}
}

\end{document}